\documentclass[final,leqno]{siamltex}

\usepackage{graphicx}
\usepackage{dcolumn}

\usepackage{amsmath,amssymb,amstext,amscd,mathrsfs,eucal,bm}
\usepackage{subfigure}
\usepackage{cases}
\usepackage{epstopdf}
\usepackage{epic,eepic,graphics,pspicture,overpic}

\usepackage{tabularx}

\usepackage{mathtools,bbm}
\usepackage{color}
\usepackage{tikz}
\usepackage{ctable}

\usepackage{tkz-euclide} 
\usetkzobj{angles} 

\usepackage{lineno}
\newcommand*\patchAmsMathEnvironmentForLineno[1]{%
  \expandafter\let\csname old#1\expandafter\endcsname\csname #1\endcsname
  \expandafter\let\csname oldend#1\expandafter\endcsname\csname end#1\endcsname
  \renewenvironment{#1}%
     {\linenomath\csname old#1\endcsname}%
     {\csname oldend#1\endcsname\endlinenomath}}%
\newcommand*\patchBothAmsMathEnvironmentsForLineno[1]{%
  \patchAmsMathEnvironmentForLineno{#1}%
  \patchAmsMathEnvironmentForLineno{#1*}}%
\AtBeginDocument{%
\patchBothAmsMathEnvironmentsForLineno{equation}%
\patchBothAmsMathEnvironmentsForLineno{align}%
\patchBothAmsMathEnvironmentsForLineno{flalign}%
\patchBothAmsMathEnvironmentsForLineno{alignat}%
\patchBothAmsMathEnvironmentsForLineno{gather}%
\patchBothAmsMathEnvironmentsForLineno{multline}%
}

\newcounter{subeq}

\DeclareMathAlphabet\mathbfcal{OMS}{cmsy}{b}{n}

\def\e{\mathrm{e}}

\def\d{\mathrm{d}}

\def\beq{\begin{equation}}
\def\eeq{\end{equation}}

\def\bx{\boldsymbol{x}}
\def\bX{\boldsymbol{X}}

\def\bu{\boldsymbol{u}}

\def\bX{\boldsymbol{X}}

\def\Da{\mathrm{Da}}
\def\Pe{\mathrm{Pe}}

\newcommand{\FK}{\textnormal{\tiny \textsc{FK}}}
\def\cfk{c_\FK}

\newcommand{\G}{\textnormal{\tiny \textsc{G}}}
\def\cg{c_\G}

\newcommand{\+}{\textnormal{\tiny \textsc{+}}}

\def\tfk{\tau_\FK}
\def\tg{\tau_\G}

\newcommand{\bs}[1]{\boldsymbol{#1}}

\def\XXint#1#2#3{{\setbox0=\hbox{$#1{#2#3}{\int}$}
\vcenter{\hbox{$#2#3$}}\kern-.5\wd0}}

		\def\beq{\begin{equation}}
		\def\eeq{\end{equation}}
		\newcommand{\eref}[1]{(\ref{eqn:#1})}
		\newcommand{\elab}[1]{\label{eqn:#1}}
		
		\newcommand{\fref}[1]{\ref{fig:#1}}
		\newcommand{\flab}[1]{\label{fig:#1}}
		
		\newcommand{\sref}[1]{\ref{sec:#1}}
  		\newcommand{\slab}[1]{\label{sec:#1}}

\def\strutdepth{\dp\strutbox}
\def\nw#1{\strut\vadjust{\kern-\strutdepth\vtop to0pt{\vss\hbox to\hsize {\hskip\hsize\hskip5pt$\leftarrow$\hss\strut}}}{\em \textcolor{blue}{#1}}}

\title{Chemical front propagation in periodic flows: \\ FKPP vs G}

\author{Alexandra Tzella\thanks{School of Mathematics, University of Birmingham, Edgbaston, Birmingham, B15 2TT, United Kingdom (a.tzella@bham.ac.uk).}
        \and Jacques Vanneste\thanks{School of Mathematics and Maxwell Institute for Mathematical Sciences, University of Edinburgh, King's Buildings,
  Edinburgh EH9 3FD, United Kingdom (J.Vanneste@ed.ac.uk).}}
  
\date{\today}

\begin{document}

\maketitle

\begin{abstract}
We investigate the influence of steady periodic flows on the 
propagation of chemical  fronts in an  infinite channel domain.  
We focus on the sharp  front  arising in Fisher--Kolmogorov--Petrovskii--Piskunov (FKPP) type models  in the limit of small  molecular diffusivity and fast reaction (large P\'eclet and Damk\"ohler numbers, $\Pe$ and $\Da$) 
and on its heuristic approximation by the  G equation.
We introduce a variational formulation that expresses the two front speeds in terms of periodic trajectories  minimizing the time of travel across the period of the flow, under a constraint that differs between the FKPP and G equations. 
This formulation makes it plain that the FKPP front speed is greater  than or equal to the G equation front speed. 
We study the two front speeds for  a class of cellular vortex flows used in  experiments.  
Using a numerical implementation of the variational formulation, 
we show that the differences between the two front speeds 
are modest for a broad range of parameters. However, large differences appear when a strong mean flow opposes front propagation; in particular, we identify a range of parameters for which FKPP fronts can propagate against the flow while G fronts cannot. 
We verify our computations against closed-form expressions derived for $\Da\ll \Pe$ and for  $\Da\gg \Pe$.

\end{abstract}


\pagestyle{myheadings}
\thispagestyle{plain}
\markboth{A. TZELLA AND J. VANNESTE}{FRONT PROPAGATION: FKPP VS G}

\section{Introduction}

A classical model for the concentration $\theta(\bx,t)$ of spreading reacting chemicals is the FKPP, or FK for short, equation 
named after the classical work by Fisher \cite{Fisher1937} and Kolmogorov, Petrovskii and Piskunov \cite{Kolmogorov_etal1937} based on logistic growth 
and diffusion. 
Numerous environmental and engineering applications, from the dynamics of ocean plankton to combustion \cite{Tel_etal2005,NeufeldHernandezGarcia2009},
 motivate 
its extension to include the effect of an incompressible background steady flow $\bu(x,y)=(u,v)$. 
The FK equation considered here then takes the non-dimensional form  
\beq \elab{FK}
\partial_t \theta + \bu \cdot \nabla \theta = \Pe^{-1} \, \Delta \theta + \Da \, r(\theta). \tag{FK}
\eeq 
The reaction term $r(\theta)=\theta(1-\theta)$ or, more generally, any function $r(\theta)$ that 
satisfies $r(0)=r(1)=0$ with $r(\theta)>0$ for $\theta \in(0,1)$, $r(\theta)<0$ for $\theta \notin[0,1]$ and $r'(0)=\sup_{0<\theta<1} r(\theta)/\theta=1$.  
The non-dimensional parameters are the  P\'eclet and Damk\"ohler numbers
\beq
\Pe=VL/\kappa\quad\textrm{and}\quad\Da=L/(V \tau),
\eeq
where $V$ and $L$ are the characteristic  speed and lengthscale of the flow, $\kappa$ the molecular diffusivity, and $\tau$  the reaction time. 
 Motivated by experiments, we  focus on two-dimensional channel domains 
 with parallel, impenetrable walls where $v=\partial_y\theta=0$ 
and take the front-like initial and boundary conditions  
\[ 
\theta(x,y,0)=\mathbbm{1}_{x\le 0}, \quad 
\theta\to 1 \quad\text{as $x\to-\infty$},
\quad
\theta \to 0 \quad\text{as $x\to\infty$},
\]
where $\mathbbm{1}$ denotes the indicator function.
In the absence of advection, \eref{FK} admits 
front solutions that propagate from the left to the right of the channel at the non-dimensional `bare' speed
\beq \elab{bare}
c_0 = 2 \sqrt{\Da/\Pe}
\eeq
corresponding to the dimensional speed $c_0^*=c_0 V$.
When the flow $\bu(x,y)$ 
is spatially periodic,
front solutions persist as pulsating fronts \cite{Xin1993,Xin2000,BerestyckiHamel2002}, 
changing periodically in time  as they travel at a speed $\cfk$, 
so that 
\beq
\theta\left(x+2\pi,y,t+{2\pi}/{\cfk}\right)=\theta(x,y,t),
\eeq
where $2\pi$ is  the spatial period of the flow. 

When reaction dominates over diffusion, i.e.\ when
\beq \elab{asymp}
\Pe\, \Da \gg 1, 
\eeq
the front interface is sharp and can be approximated by a single curve (in 2D as assumed here)  where all the reaction takes place.
A distinguished regime then arises for 
\beq
\quad \Da/\Pe = c_0^2/4 = O(1),
\eeq
when advection and reaction--diffusion both contribute to the front propagation at the same order.
In these conditions, a heuristic model is often used in place of \eref{FK}. In this model, the front is the zero-level curve    
$\theta(x,y,t)=0$, say, where $\theta(x,y,t)$  
satisfies the Hamilton--Jacobi equation \beq\elab{G}
\partial_t \theta + \bu \cdot \nabla \theta = c_0 |\nabla \theta|, \tag{G}
\eeq
termed G equation \cite{Williams1985} (see also  \cite{Sethian1985,Kerstein_etal1988}).
This model is popular in the combustion science literature (e.g.\ \cite{Peters} and references therein).
For $\bu=\bm{0}$, the front speed predicted by \eref{G} is obviously $c_0$, matching the speed predicted by \eref{FK}. 
For spatially periodic $\bu\not=\bm{0}$, \eref{G} predicts pulsating front solutions propagating with a speed $\cg$  
that  in general differs from $\cfk$ \cite{XinYu2010,Cardaliaguet_etal2011}. The relation between the two speeds $\cfk$ and $\cg$ 
(with dimensional equivalents $\cfk^*=\cfk V$ and $\cg^*=\cg V$) is the subject of this paper. 

Majda and Souganidis \cite{MajdaSouganidis1994}  showed that 
in the limit \eref{asymp} 
the leading-order $\cfk$ can be deduced 
from the long-time solution of a certain Hamilton--Jacobi equation. 
This long-time solution is obtained by applying the asymptotic procedure of homogenisation \cite{Lions_etal,Evans_etal1989} which
exploits spatial scale separation
to express $\cfk$ in terms of  the eigenvalue of 
a   nonlinear  cell problem posed over a single period of the flow.
A similar procedure can be applied to \eref{G}, leading to a different nonlinear eigenvalue 
cell problem  
for $\cg$.
The two nonlinear cell problems are significantly simplified for the special case of shear  flows \cite{Embid_etal1995,XinYu2013}. 
For more general flows  and arbitrary $c_0$ 
explicit analytical expressions  are not  available and the two  cell problems  need to be solved numerically.
However, these computations can be rather challenging (see e.g. \cite{KhouiderBourlioux2002} for the nonlinear cell problem related to  $\cfk$).
 Analytic work has focused on the strong-flow limit corresponding to $c_0 \to 0$ \cite{SmailyKirsch2011,XinYu2013,XinYu2014}.

In this paper, we rely on the variational representation of the 
two front speeds $\cfk$ and $\cg$. For \eref{FK}, this approach  was introduced by  Freidlin and collaborators  (see \cite[Ch.~10]{FreidlinWentzell1984}, \cite[Ch.~6]{Freidlin1985} and \cite{FreidlinSowers1999}) to 
establish an expression for  $\cfk$ in terms of a single  trajectory that minimises an action functional. This   
was subsequently  exploited in \cite{TzellaVanneste2014} to obtain explicit results for cellular flows by carrying out a minimisation over periodic trajectories. 
For \eref{G}, Fermat's principle in a moving medium  determines $\cg$. 
The variational formulations enable us to express $\cfk$ and $\cg$ in terms  
of periodic trajectories 
$\bX(\tau)=\bX(0)+(2\pi,0)$ that minimise the time of travel 
$\tau$ across the period of the flow,  under a  constraint that differs between \eref{FK} and \eref{G}. 
  In both cases, the constraint 
involves the    
difference between the velocity of the minimising trajectory and the velocity of the flow. 
For \eref{FK}  the constraint 
is integral, in terms of the   $L^2$-norm,  
given by 
\beq \elab{consFK}
\tau^{-1} \int_0^\tau |\dot \bX(t) - \bu(\bX(t))|^2 \, \d t = c_0^2
\eeq
while for  \eref{G} the constraint is  pointwise and
given by
\beq \elab{consG}
|\dot \bX(t) - \bu(\bX(t))|^2 =c_0^2, 
\eeq
for all $t\in[0,\tau]$.
 These formulations allow us
 to  understand  the difference between $\cfk$ and $\cg$, 
  to  immediately deduce that   $\cfk\geq \cg$ (already established by \cite{XinYu2013} using a different approach) 
 and to compute 
  $\cfk$ and $\cg$ for a large class of steady, periodic 
 $\bu$.

	\begin{figure*}
		\begin{center}
		\begin{overpic}[width=0.9\linewidth]{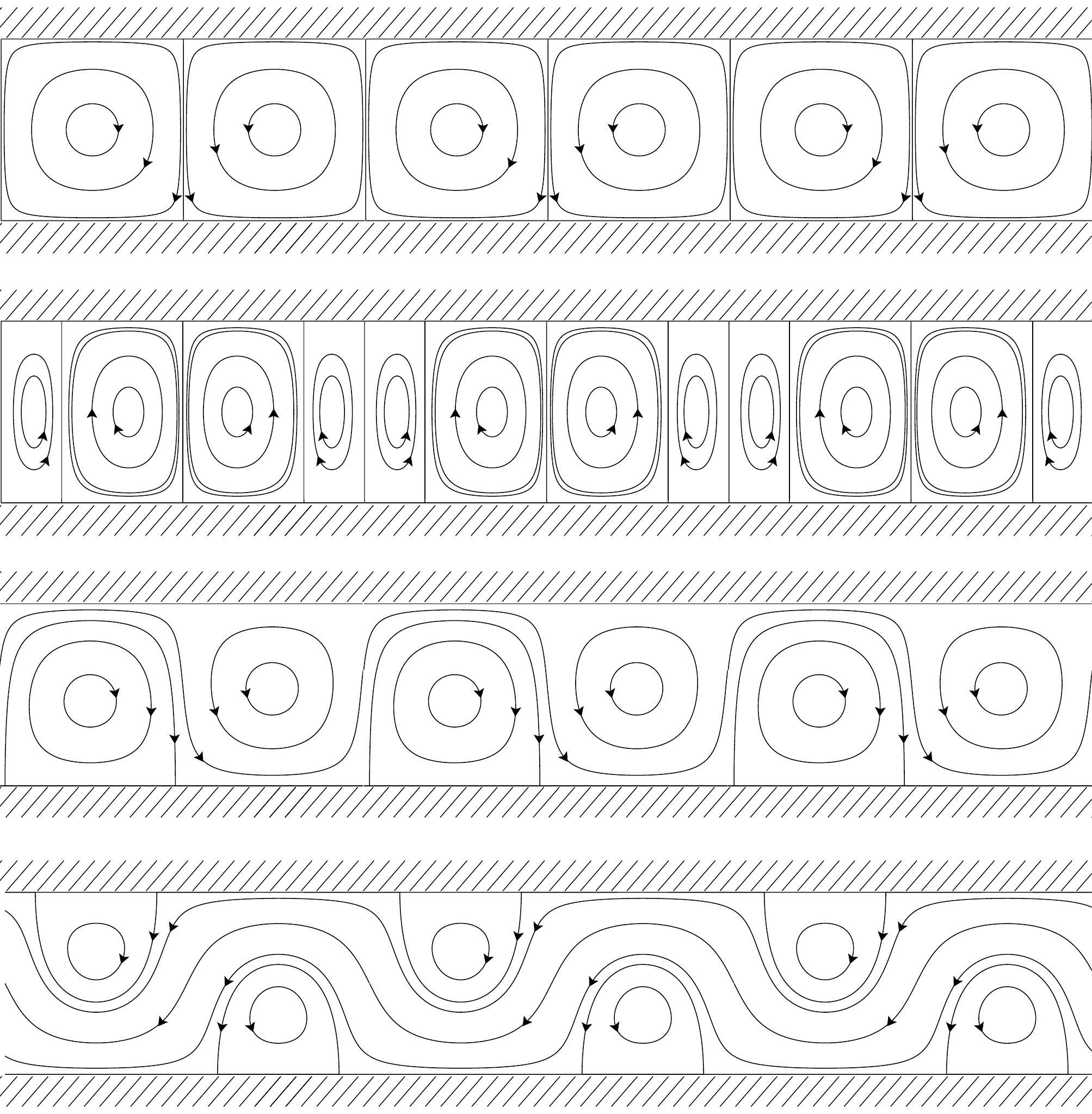}
		\put(-4.5,96){(a)}
		\put(-4.5,71){(b)}
			\put(-4.5,46){(c)}
			\put(-4.5,21){(d)}
					\end{overpic}
		\end{center}
	\caption{
	Streamlines for the cellular  flow with streamfunction \eref{streamfunction} for   (a) $U=0$, $A=0$,  (b) $U=0$, $A=1$, (c) $U=0.1$, $A=0$ and (d) $U=-0.5$, $A=0$.  
	 For   $U=0$, all streamlines are closed. When $U \not= 0$, there is a channel of open streamlines.}
	\flab{streamfnarray}
	\end{figure*}
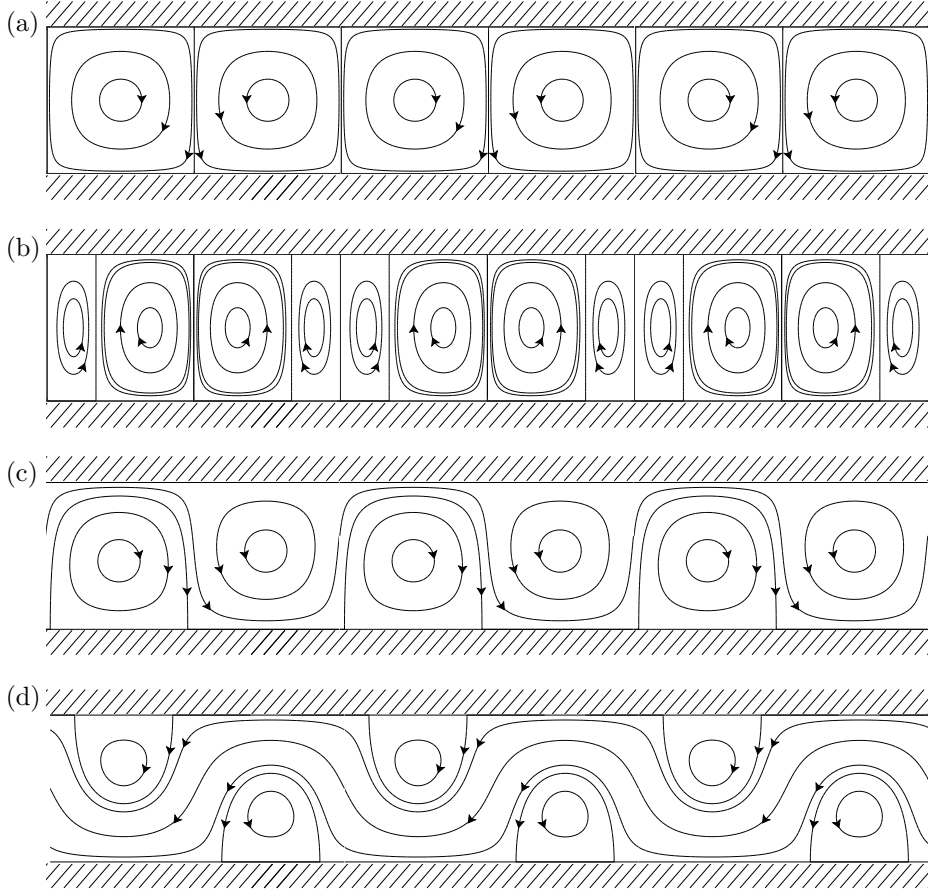

 We  begin with  the simple case of shear flows $\bu=(u(y),0)$ before examining in detail 
 a two-parameter family of periodic  cellular  flows, given by    
  $\bu=(-\partial_y \psi,\partial_x \psi)$
 with streamfunction 
 \beq \elab{streamfunction}
 \psi = -Uy-\left(\sin x +A\sin(2x)\right)\sin y. 
 \eeq
 This is used  as a testbed in numerous experimental studies of advection--diffusion--reaction 
 (e.g., \cite{PocheauHarambat2008,SchwartzSolomon2008,BargteilSolomon2012, Megson_etal2015, Mahoney_etal2015}).
 The classic   cellular   flow  introduced in \cite{Roberts1972}  corresponds to   a zero mean velocity $U=0$  and to $A=0$.  
 When confined  between  
 walls at $y=0$ and $\pi$, 
 this flow consists of a one-dimensional infinite array of 
 periodic cells composed of two vortices of opposite circulation.
 These   vortices are bounded by the separatrix  streamline $\psi=0$  that connects a network of hyperbolic stagnation points    
 (see  Fig.\ \fref{streamfnarray}(a)). 
 All streamlines remain closed when  $A>0$ and $U=0$ but the symmetry $(x,y) \mapsto (x+\pi,\pi-y)$ is broken.
 For $A>1/2$, the number of hyperbolic stagnation points doubles and the periodic cell consists of four  vortices 
 rotating in alternatively clockwise  and anticlockwise directions 
 (see  Fig.\ \fref{streamfnarray}(b)).  
 The  topology of the streamlines  changes drastically for a non-zero mean  velocity $U \not= 0$:
 an open channel, bounded by the separatrices $\psi=0$ and $\psi=-U\pi$, traverses the domain, splitting apart the row of closed vortices. 
 As the value of $|U|$ increases, the width of the open channel increases (see Fig.\ \fref{streamfnarray}(c) for $U>0$ and Fig.\ \fref{streamfnarray}(d) for $U<0$).   For $|U|$ large enough, the hyperbolic stagnation points and closed streamlines disappear.

Our aim  is to  determine  the effect of  flow structures on the value of the two front speeds $\cfk$ and $\cg$ and on their difference.  
To achieve this, we develop and implement a highly accurate numerical method  
that is based on the  efficient discretisation of a pair of variational principles that we obtain.   
Computations of  the two front speeds are 
complemented by  a set of explicit expressions derived 
by formal asymptotics methods in the limit of small and large values of $c_0$ and various   values of $A$ and $U$. 
Table \ref{tbl1} summarises the expressions for the basic cellular flow  for which $A=U=0$. 
These  are  in agreement with the  rigorous bounds developed in  \cite{XinYu2013} for small $c_0$ (see also \cite{Abel_etal2002,Cencini_etal2003}).

The paper is organised as follows. In section \sref{FrontSpeed},  we provide a 
brief derivation 
of the two   nonlinear cell problems that determine $\cfk$ and $\cg$. 
In section \sref{variational}, we introduce the alternative characterisation  in the form of a pair of variational principles with constraints \eref{consFK}--\eref{consG}. 
The two principles greatly simplify for shear flows in which case $\cfk=\cg$.  
Section \sref{results} is devoted to  flows with streamfunction \eref{streamfunction}.
The numerical scheme employed for the computations  is described in the Appendix.
The paper ends with a discussion in section \sref{conc}.

\begin{table}
\caption{
Asymptotic expressions for the  front speed of \eref{FK}  and \eref{G} in  the basic cellular flow 
(\eref{streamfunction}
with   $A=U=0$) for small and large  `bare' speed $c_0=2\sqrt{\Da/\Pe}$. 
The difference between the two front speeds is asymptotically  small in both limits (see Sec. \sref{cellular} for  details). All variables are non-dimensional. 
}
\begin{center} \footnotesize
\begin{tabular}{cccc} 
\specialrule{.1em}{.5em}{.5em} 
 equation &   front speed  $\sim$  & range of validity \\
\specialrule{.1em}{.5em}{.05em} 
\\
\eref{FK}          &    $\pi/\mathrm{W}_p(32c_0^{-2})$             &  $c_0\ll 1$
\\[5pt]
           &    $  c_0(1+3 c_0^{-2}/4-105 c_0^{-4}/64)$             &  $c_0\gg 1$
\\
\specialrule{.05em}{1em}{0em} 
\\
\eref{G}        &    $-\pi/(2\log(\pi c_0/8))$  & $c_0\ll 1$
\\[5pt]
         &    $ c_0(1+3 c_0^{-2}/4-109 c_0^{-4}/64)$  & $c_0\gg 1$
\\[5pt]
\specialrule{.1em}{.05em}{.05em} 
\\
\end{tabular}
\end{center} 
\label{tbl1}
\end{table}

\section{Front speed}\slab{FrontSpeed}
\subsection{Equation \eref{FK}}
G\"artner and Freidlin \cite{GartnerFreidlin1979} showed that 
for initial conditions  sufficiently close to a step function,
the  speed of the front  associated with  \eref{FK} can be deduced by the long-time behaviour of the solution near the front's leading edge. 
There $0< \theta\ll 1$ and $r(\theta)\approx r'(0)\theta=\theta$ so that \eref{FK} becomes
\beq \elab{FKlin}
\partial_t \theta + \bu\cdot \nabla \theta = \Pe^{-1} \, \Delta \theta + \Da \, \theta.  
\eeq
For $\Pe \gg 1$ and $\Da/\Pe = c_0^2/4 = O(1)$, the solution can be sought in the WKBJ (Wentzel--Kramers--Brillouin--Jeffreys) or geometric-optics form
\beq\elab{FKlinsol}
\theta(\bx,t) \asymp \e^{-\Pe \,  \mathscr{I}(\bx,t,c_0) }.
\eeq
Collecting the terms with the same powers in  $\Pe$, we find that at leading order  $\mathscr{I}(\bx,t,c_0)$ satisfies the Hamilton--Jacobi equation
\beq\elab{I1}
 \partial_t \mathscr{I} + \mathscr{H}_{\FK}(\nabla \mathscr{I},\bx,c_0)  = 0 \quad \textrm{with} \ \ \mathscr{H}_{\FK}(\bs{p},\bs{x},c_0) = |\bs{p}|^2 + \bu(\bx) \cdot \bs{p} + c_0^2/4
\eeq
 the Hamiltonian. 
 The step-function initial conditions correspond to
 $\mathscr{I}(\bx,0,c_0)=0$ for $x\leq 0$ and $\mathscr{I}(\bx,0,c_0)=\infty$ for $x>0$, and the boundary conditions to $\partial_y\mathscr{I}(\bx,t,c_0)=0$ at  $y=0$, $\pi$.
The front  is then  identified as the location where \eref{FKlinsol}  neither grows nor decays exponentially with time. It is therefore  the level curve
 \beq\elab{I2b}
   \mathscr{I}(\bx,t,c_0)=0.
 \eeq

In the long-time limit, the solution to \eref{I1}
converges to that of the homogenised 
Hamilton--Jacobi equation  
 \beq\elab{I5}
 \partial_t \bar{\mathscr{I}} + \bar{\mathscr{H}}_{\FK}(\partial_x \bar{\mathscr{I}},c_0) = 0.
 \eeq
The effective Hamiltonian, $\bar{\mathscr{H}}_{\FK}$, may be derived from 
 a nonlinear
eigenvalue problem, obtained 
by writing the solution to \eref{I1}   
as the multiscale    expansion 
 \beq\elab{I3}
 \mathscr{I}(\bx,t,c_0)=t \left(\mathscr{G}(c,c_0)+t^{-1}\phi(\bx,c,c_0)+O(t^{-2})\right),
 \ \ \textrm{where} \ \ t\gg 1 \ \ \text{and} \ \ c=x/t=O(1).
 \eeq
Here $c$ is the slow variable describing the speed of a moving frame of reference and  $\bx$ is the fast variable.   
We emphasise
the particular form of \eref{I3}, with a leading-order term 
  that is   independent of  
   $\bx$ and involves 
$\mathscr{G}(c,c_0)$ that depends  on $c$ only.\footnote{Note that $\mathscr{G}(c,c_0)$  
may be interpreted as 
the Freidlin--Wentzell (small-noise, large-$\Pe$) large-deviation rate function 
for the position of fluid particles that have been displaced
by advection and diffusion  to a distance $ct$ in a time $t\gg 1$ 
 (see \cite{FreidlinWentzell1984},\cite[Ch.~6]{Freidlin1985} and \cite{FreidlinSowers1999} for rigorous treatments). } 
The next order involves $\phi(\bx,c,c_0)$ where $\phi(x+2\pi,y,c,c_0)=\phi(x,y,c,c_0)$ 
while the boundary conditions at $y=0$, $\pi$ imply that there, $\partial_y \phi=0$. 
Substituting \eref{I3} into \eref{I1} and equating powers of $t^{-1}$ yields at leading order
$O(1)$ 
the  nonlinear eigenvalue problem 
\beq\elab{I4}
\mathscr{H}_{\FK}((p,0)+\nabla\phi,\bs{x},c_0)=\bar{\mathscr{H}}_{\FK}(p,c_0),
\ \ \text{where} \ \ 
p=\mathscr{G}'(c,c_0),
\eeq
with the prime denoting derivative with respect to the first argument, can be  treated as a parameter and 
\beq \elab{Hbar}
\bar{\mathscr{H}}_{\FK}(p,c_0)=c\,  \mathscr{G}'(c,c_0)-\mathscr{G}(c,c_0),
\eeq
is the eigenvalue.  
It can be shown that $\bar{\mathscr{H}}_{\FK}(p,c_0)$ is unique, non-negative, real and convex in $p$ 
(see \cite{Lions_etal,Evans1992} for  proofs) 
and therefore $\bar{\mathscr{H}}_{\FK}(p,c_0)$  and $\mathscr{G}(c,c_0)$ 
are related  via a Legendre transform  
\beq\elab{leg} \mathscr{G}(c,c_0)=\sup_{p}(p\,c-\bar{\mathscr{H}}_{\FK}(p,c_0))\quad\text{and}\quad \bar{\mathscr{H}}_{\FK}(p,c_0)=\sup_{c}(p\,c-\mathscr{G}(c,c_0)).
\eeq

Combining \eref{I2b} with \eref{I3} 
gives the front speed $\cfk$  as the solution of 
\beq \elab{Gc02}
\mathscr{G}(\cfk,c_0)=0,
\eeq
with $\cfk>0$   corresponding to \eref{FK} fronts that propagate from left to  right.  
Using  \eref{leg} it can be expressed explicitly in terms of the effective Hamiltonian $\bar{\mathscr{H}}(p)$  as
\beq\elab{speed1}
\cfk=\inf_p \frac{1}{p} \bar{\mathscr{H}}_{\FK}(p,c_0),
 \eeq
an expression first obtained in \cite{MajdaSouganidis1994}.  

\subsection{Equation \eref{G}} The long-time solution to equation \eref{G} can be treated similarly.
It satisfies the homogenised Hamilton--Jacobi equation
 \beq\elab{I6}
 \partial_t \bar{\theta} + \bar{\mathscr{H}}_{\G}(\partial_x \bar{\theta},c_0) = 0, 
 \eeq
with an effective Hamiltonian $\bar{\mathscr{H}}_{\G}$ found as eigenvalue of the nonlinear cell problem 
\beq\elab{nonlinearcellG}
\mathscr{H}_{\G}((p,0)+\nabla\phi,\bs{x},c_0)=\bar{\mathscr{H}}_{\G}(p,c_0), 
\eeq
where
\beq\elab{HamiltonianG}
\mathscr{H}_{\G}(\bs{p},\bx,c_0)=\bu(\bs{x}) \cdot\bs{p}-c_0|\bs{p}|.
\eeq
Note that the nonlinearity  $|\bs{p}|^2$ in $\mathscr{H}_{\FK}$ is replaced here by $|\bs{p}|$. 
Nevertheless, $\bar{\mathscr{H}}_{\G}$ is unique and convex 
 (details and proofs can be found in  \cite{XinYu2010,Cardaliaguet_etal2013}). 
The solution of \eref{I6} is then $\bar{\theta}=t\mathscr{F}(c,c_0)$ where $\mathscr{F}(c,c_0)$ and $\bar{\mathscr{H}}_{\G}(p,c_0)$  and are   related via a Legendre transform analogous to \eref{leg}.  
Since the front corresponds to $\theta(\bx,t)=0$,   
in the long-time limit, the  speed $c_\G$  of right-propagating \eref{G} fronts is found as the positive solution of $\mathscr{F}(\cg,c_0)=0$ or, equivalently, as 
\beq\elab{speed2}
\cg=\inf_p  \frac{1}{p}\bar{\mathscr{H}}_{\G}(p,c_0).
\eeq

\vspace{1cm}

We now obtain alternative formulations to \eref{speed1} and \eref{speed2}  that shed light on  the difference between the two speeds, are amenable to straightforward numerical computations, and yield explicit expressions in asymptotic limits.

\section{Variational principles}\slab{variational}
\subsection{Equation \eref{FK}} It is well known (see e.g. \cite{Evans}) that the  solution to \eref{I1} may be written as a variational principle involving an 
action functional associated with the Lagrangian 
\beq\elab{Lag}
\mathscr{L}(\dot{\bX},\bX)=\frac{1}{4}| \dot{\bX} - \bu(\bX)|^2
\eeq
that is dual to  the Hamiltonian $\mathscr{H}_{\FK}$ in  \eref{I1}.
For $x>0$ the solution  is  given by 
\begin{align}\elab{I2}
\mathscr{I}(\bx,T,c_0)& = \frac{1}{4} \left( \inf_{\bX(\cdot)} \int_0^T | \dot{\bX}(t) - \bu(\bX(t))|^2 \, \d t -c_0^2 T \right),\\
&\qquad\qquad\qquad\qquad\qquad  \textrm{subject to} \   \bX(0)=(0,\cdot),  \ \ \bX(T)= \bx, 
\end{align}
where $\bX(\cdot)$ represents a family of smooth trajectories with  $Y(\cdot)\in [0,\pi]$. 
From \eref{I3}  we have  
\beq\elab{G1}
\mathscr{G}(c,c_0)=  
\lim_{T\to\infty} \frac{\mathscr{I}((c T,y),T,c_0)}{T},
\eeq
where the dependence on the specific value of $y$ drops out
(e.g.\ \cite{Piatnitski1998}). Together with \eref{I2} 
this determines the function $\mathscr{G}(c,c_0)$.

Expression \eref{G1} can be  simplified using the spatial periodicity of the background velocity $\bu$ \cite{TzellaVanneste2014}. 
Assuming that the minimising trajectory inherits 
inherits the same spatial periodicity,
we take 
$T = n \tau $ with $\tau=2\pi/c$ and $n \gg 1$ to reduce  \eref{G1}  to
\begin{align}\elab{Glong}
\mathscr{G}(c,c_0) & =  \frac{1}{4} \left(\frac{1}{\tau} \inf_{\bX(\cdot)} 
\int_0^{\tau} | \dot{\bX}(t) - \bu(\bX(t))|^2 \, \d t - c_0^2 \right), \\
 &\qquad\qquad\qquad\qquad\qquad \textrm{subject to} \  \bX(\tau)=\bX(0)+(2\pi,0). \nonumber	 
\end{align}
 Expression \eref{Glong} provides a direct way to compute   the minimising trajectory and,  from  \eref{Gc02}, the corresponding 
 front speed $\cfk$, both numerically and in asymptotic limits. Such computations were carried out in \cite{TzellaVanneste2014} 
 for the specific case of the   cellular flow with closed streamlines that we consider further in Section \sref{results}. 
These computations were validated   
against 
the numerical evaluation of $\cfk$ for finite P\'eclet and  Damk\"ohler
numbers 
obtained from    an 
advection--diffusion eigenvalue problem
 and   direct numerical simulations of \eref{FK} with $r(\theta)=\theta(1-\theta)$.

We now obtain an alternative variational characterisation of $\cfk$.
Since $\cfk$ satisfies $\mathscr{G}(\cfk,c_0)=0$, it can be written as extremum 
of the function    
\beq 
  S(\lambda)
 =\sup_{\tau}\frac{2\pi}{\tau} - \lambda \mathscr{G}\left(\frac{2\pi}{\tau},c_0\right), \elab{FKVar3}
\eeq
for arbitrary variations of the Lagrange multiplier $\lambda$. Here we use that $\mathscr{G}$ is convex in $c$, so that a single $\tau=\tfk$ satisfies the constraint $\mathscr{G}(2\pi/\tau,c_0)=0$ enforced by $\lambda$. Using \eref{Glong} and redefining $\lambda$ to absorb a factor $1/4$, we can rewrite this as 
\begin{align}
 S(\lambda) &= 
\sup_{\tau}\sup_{\bX(\cdot)} \left(
\frac{2\pi}{\tau}-\lambda \left(\frac{1}{\tau}\int_0^\tau |\dot \bX(t) - \bu(\bX(t))|^2  \d t -c_0^2\right) \right), \elab{FKVar1}
\\
 &\qquad\qquad\qquad\qquad\qquad
 \  \textrm{subject to} \  \  \bX(\tau)=\bX(0)+(2\pi,0).   
\end{align} 
This can be interpreted as the maximisation of $2\pi/\tau$ under a constraint enforced by the Lagrange multiplier $\lambda$. 
Therefore, the front speed predicted by  \eref{FK} for $\Pe, \, \Da \gg 1$, $c_0=O(1)$ is given as
\begin{align} \elab{cfkvar}
	\cfk  = \frac{2 \pi}{\tfk},  \ \  \textrm{where} \ \ \tfk = \inf_{\bX(\cdot)} \tau,   \  \ &\textrm{subject to} \  \  \bX(\tau)=\bX(0)+(2\pi,0)   
	                 \\
	                 &\textrm{and}   \   \frac{1}{\tau} \int_0^{\tau} |\dot \bX(t) - \bu(\bX(t))|^2 \d t= c_0^2.  \nonumber
\end{align}
This variational characterisation expresses $\cfk$ as the maximum mean velocity achievable by periodic trajectories that are constrained to depart from passive-particle trajectories in a prescribed way. 

\subsection{Equation \eref{G}}
An analogous variational characterisation describes the front speed associated with  \eref{G}. 
Taking the same initial conditions as for  \eref{FK}, 
the  front propagates from its initial location at $\bX(0)=(0,\cdot)$ along trajectories $\bX(t)$ that obey Fermat's principle in a moving medium (e.g.\ \cite{CourantHilbert1991}, Vol.\ 1, Sec.\ IV.1). Thus the front reaches location $\bx$ after
a travel time
\begin{align} \elab{cgvar0}
	\mathscr{T}(\bx,c_0) = \inf_{\bX(\cdot)} T \  \ &\textrm{with} \  \ \bX(0)=(0,\cdot),\ \bX(T)=\bx,   
	                 \\
					  &\textrm{subject to}\  |\dot \bX(t) - \bu(\bX(t))|^2 = c_0^2 \ \textrm{for} \ t \in [0,T],  \nonumber	 
\end{align}
where  again we assume that $\bX(\cdot)$ represents a family of smooth trajectories with  $Y(\cdot)\in [0,\pi]$.
In the long-time limit,  $x$ is large and the front moves at a constant speed given by  
\beq\elab{cgvar1}
\cg=
\lim_{x\to\infty} \frac{x}{\mathscr{T}((x,y),c_0)},
\eeq
where once more the dependence on $y$ drops out.  
This characterisation 
is significantly simplified if we  apply the same strategy as before and assume   that the minimising trajectory is periodic. Taking $T=n\tau$ with $n\gg 1$, we obtain that  
\begin{align} \elab{cgvar}
	\cg  = \frac{2\pi}{\tg}, \ \textrm{where} \ \tg = \inf_{\bX(\cdot)} \tau, \  \ &\textrm{subject to} \  \  \bX(\tau)=\bX(0)+(2\pi,0)  
	                 \\
					  &\textrm{and}\  |\dot \bX(t) - \bu(\bX(t))|^2 = c_0^2 \ \textrm{for} \ t \in [0,\tau].  \nonumber	 
\end{align}
This characterisation of the front speed for \eref{G} closely parallels the characterisation \eref{cfkvar} of the front speed for \eref{FK}.

For practical computations, it is convenient to rewrite \eref{cgvar} taking $x$ as the independent variable,
using 
\beq
\frac{dt}{dx}=T'(x),\ \ \textrm{with} \ \ T(0)=0,
\eeq
where $T(x)$ denotes the time it takes to reach the point $(x,Y(x))$. 
The minimal travel time over a spatial period   is then expressed as
\begin{align} \elab{cgvar2}
	\cg  = \frac{2\pi}{\tg}, \ \textrm{where} \ \tg &=  \inf_{T(\cdot),Y(\cdot)} \, \int_0^{2\pi} T'(x) \d x \  \ 
	\textrm{subject to} \  \   Y(2\pi)=Y(0)
	                 \\
					  &\textrm{and}\  |T'(x)^{-1}(1,Y'(x)) - \bu(x,Y(x))|^2 = c_0^2 \ \textrm{for} \ x \in [0,2\pi],  \nonumber	 
\end{align}
and   $Y(\cdot)$, $T(\cdot)$ are taken to be smooth.

\subsection{Comparison} We now compare the two variational characterisations \eref{cfkvar} and \eref{cgvar} for the \eref{FK} and \eref{G} equations. In both the front speeds are expressed in terms of the travel times $\tfk$ and $\tg$ which are 
determined by the periodic trajectories that traverse a spatial period of the flow in the least  time. 
The only difference    is that the pointwise constraint on the relative velocity in \eref{cgvar} is replaced by a slacker, time-averaged constraint in \eref{cfkvar}. An immediate consequence is that 
\beq \elab{ineq}
\cfk \ge \cg,
\eeq
The same result was obtained in \cite{XinYu2013} using a min-max formulation of \eref{I4} and \eref{nonlinearcellG}.

While \eref{cfkvar} and \eref{cgvar} are useful for comparisons of this type, for numerical computations we found it convenient to use 
\eref{Glong} and \eref{cgvar2} instead. Eq.\ \eref{Glong} is useful for  \eref{FK}  when, as is the case in section \sref{results}, we are interested in computing $\cfk$ for a range of values of $c_0$: the simple dependence of $\mathscr{G}$ on $c_0$ means that the condition $\mathscr{G}(\cfk,c_0)=0$ gives an explicit variational formula for $c_0$ as a function of $\cfk$ with the endpoint condition as  sole constraint.

The variational characterisation \eref{cgvar} is also useful to establish a necessary condition for the existence of right-propagating front solutions for the \eref{G} equation. It is easy to see from the constraint in \eref{cgvar} that 
\beq\elab{stat_G}
\cg > 0\quad\text{implies}\quad c_0 > -\min_{x}\max_y u(x,y).
\eeq
For smaller $c_0$, there are no right-propagating 
\eref{G} fronts. 
From \eref{ineq} we then expect that, for a range of $c_0$, there exist right-propagating fronts for  \eref{FK}  but not for  \eref{G}. We provide explicit examples confirming this in section \ref{sec:meanflow}.

\subsubsection*{Shear flows}
It is easy   to show that for shear flows with velocity $\bu(\bx)=(u(y),0)$, $\cfk =\cg$. 
For \eref{FK}, the  Euler--Lagrange equations associated with the functional in \eref{Glong} can be written as
\beq
\dot X(t) - u(Y(t)) = A_1, \quad \tfrac{1}{2} \dot Y^2(t) + A_1 u(Y(t)) = A_2,
\eeq
where $A_1$ and $A_2$ are two constants. The minimum of the functional is then achieved when $Y(t)=Y_0$, where $Y_0$ is a constant to be determined. It follows that $\dot X(t) = \mathrm{const} = c$ as imposed by the endpoint condition. The functional then reduces to $(c-u(Y_0))^2$. Its minimum is non zero for $c>u_{\+}=\max_{y} u(y)$, the maximum  velocity in the channel, and given by $(c-u_{\+})^2$ with $Y_0=Y_{\+}$ such that $u(Y_{\+})=u_{\+}$. Thus, 
\beq \elab{Gshear}
\mathscr{G}(c,c_0)=\left((c-u_{\+})^2-c_0^2\right)/4 \ \ \ \textrm{for} \ \ c> u_{\+},
\eeq
and solving \eref{Gc02} gives the front speed
$\cfk=c_0+u_{\+}$.

On the other hand, the pointwise constraint \eref{cgvar}  of the velocity may be parameterised 
so that 
\beq
\dot X(t)= u(Y(t))+c_0\cos\Theta(t)
\quad\textrm{and}\quad
\dot Y(t)= c_0\sin\Theta(t), 
\eeq
where $\Theta(t)$ has the same period as $\bX(t)$. 
The minimum value of $\tau$ is obtained by 
maximising $\dot X(t)$. 
This is achieved for $\dot Y(t)=0$, $\Theta(t)=0$ and $Y=Y_{\+}$, i.e. for trajectories that follow the (straight) streamline associated with maximal flow velocity. We deduce that
\beq\elab{cshear}
\cfk=\cg=c_0+u_{\+}.
\eeq 
We therefore conclude that  \eref{FK} and  \eref{G} are equivalent in describing
 the long-time  speed of propagation. This was previously 
argued to be the case in \cite{Audoly_etal2000}, can be inferred from the analysis in \cite{Embid_etal1995}  and was proved in \cite{XinYu2013}. It is clear that  a right-propagating front is obtained for both \eref{FK} and \eref{G}  provided that $c_0>-u_{\+}$, and that the front is stationary for $c_0=-u_{\+}>0$.

\section{Front speeds for periodic flows}\slab{results}

For more general flows, closed-form formulas are not available.   
We use the variational problems \eref{Glong}  and \eref{cgvar2} whose solutions  
are easy to approximate numerically. 
We obtain numerical approximations by discretising trajectories, action functional 
and constraints and determining the optimal solutions by minimisation.
The numerical procedure is detailed in Appendix \sref{num}.  
We use this procedure to compute the front speeds for \eref{FK} and \eref{G} and a range of two-dimensional periodic flows. We now describe the results.

\subsection{Cellular flow}\slab{cellular}
We first compute the solutions    for the  closed cellular flow with streamfunction \eref{streamfunction} and  $U=A=0$.  
Figure \fref{trajectories_cell} shows characteristic examples of  minimising trajectories
 obtained for three different values of $c_0$. 
For large values of $c_0$, 
the periodic trajectories for \eref{FK} and \eref{G}
are close to the straight line $y=\pi/2$. In this case, the two   trajectories are practically indistinguishable. 
A larger difference is obtained for small values of $c_0$, in which case both  trajectories follow closely a streamline near the separatrix $\psi=0$. 
In all cases it is clear that the trajectories are invariant under the transformations $(x,y) \mapsto (-x,\pi-y)$ and $(x,y) \mapsto (x + \pi,\pi-y)$.

Figure \fref{speed_cell_asymptotics} shows the behaviour of the  front speeds  for \eref{FK} and \eref{G}  as a function of $c_0$.    
Clearly, there is a difference between $\cfk$ and $\cg$ which is more marked for smaller values of $c_0$.  
However, this difference is small:  \eref{G} only slightly underpredicts the front speed of \eref{FK}.  
The behaviour of $\cfk$ and $\cg$ and their difference  can be captured by explicit expressions obtained in two asymptotic limits.

\begin{figure*}
	\centerline{\includegraphics[width=0.5\linewidth]{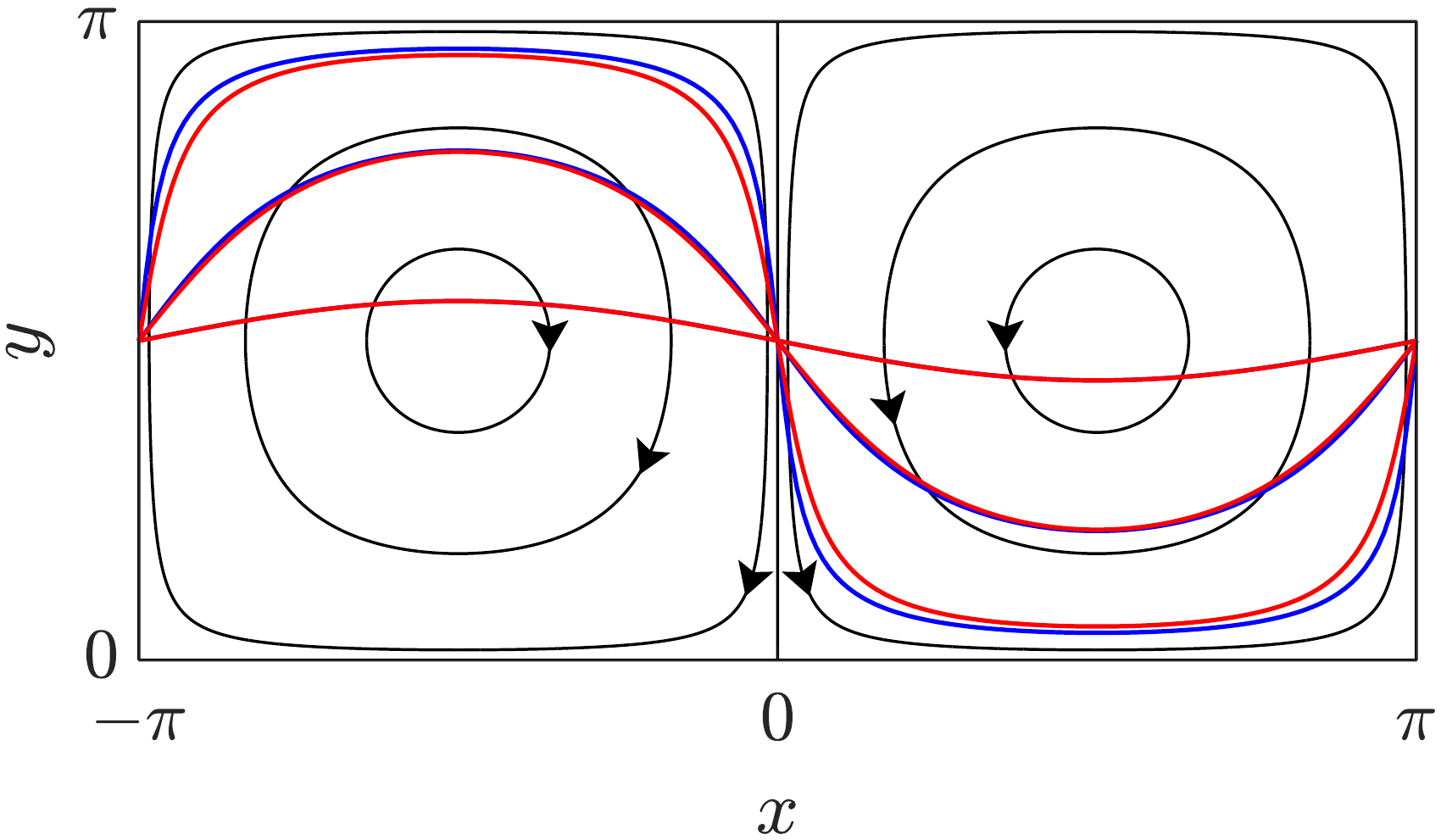}}
\caption{(Color online).  Streamlines (thin  black lines) of  the closed cellular flow with streamfunction \eref{streamfunction} and  $U=A=0$. 
and corresponding periodic trajectories for \eref{FK} (minimising \eref{cfkvar}, thick blue  lines) and   \eref{G} (minimising \eref{cgvar}, thick red  lines) 
obtained numerically for $c_0=0.1$, $c_0=1$ and $c_0=10$. The trajectories become closer to the straight line $y=\pi/2$ as $c_0$ increases 
at which point the difference between the two sets of trajectories is minimal.
 }
\flab{trajectories_cell}
\end{figure*}

\subsubsection{Small-$c_0$ asymptotics} The first asymptotic limit corresponds to $c_0\ll 1$. 
This limit has been studied  in  \cite{XinYu2013} who rigorously derived tight bounds on $\cfk$ and $\cg$.
We find an approximation to $\cfk$ 
by approximating $\mathscr{G}(c,c_0)$ in \eref{Glong} for $c \ll 1$. 
We previously found \cite{TzellaVanneste2014} that 
the minimising periodic trajectory in \eref{Glong}  
may be divided into two regions that we now describe. 
In   region I, $X(t)\ll 1$ and
therefore we may seek  a regular expansion in powers of $c$ of the form
\beq\elab{expansion1_smallc0a}
\bX(t)=(0,Y_0(t))+c(X_1(t),Y_1(t))+\cdots,
\eeq
 where, without loss of generality, we take $X(0)=0$. 
In   region II, $Y(t)\ll 1$ and so we take
\beq\elab{expansion2_smallc0a}
\bar{\bm{X}}(t)=(\bar{X}_0(t),0)+c(\bar{X}_1(t),\bar{Y}_1(t))+\cdots,
\eeq
where $\bar{X}(\tau/4)=\pi/2$ with $\tau=2\pi/c$.  
 We then exploit the  symmetries that
  characterises the streamfunction   to extend the trajectory over the whole time period $\tau$.

Substituting  \eref{expansion1_smallc0a} and \eref{expansion2_smallc0a}
into  \eref{Glong} gives a sequence of integrals corresponding to successive powers of $c$.  Minimising each yields
\begin{subequations}
	\begin{align}
	\dot{Y}_0&=-\sin Y_0,\enspace \ddot{X}_1=X_1,\enspace \dot{Y}_1= -Y_1\cos Y_0,\\ 
 	\dot{\bar X}_0&=\sin \bar{X}_0,\enspace \dot{\bar X}_1={\bar X}_1\cos X_0 ,\enspace \dot{\bar Y}_1=-\bar{Y}_1\cos\bar{X}_0.
 \end{align}
 \end{subequations}
Thus at $O(c)$ in Region II, the minimising trajectory follows exactly the streamlines. The two solutions can be  matched   in their common region of validity, given by $X(t), Y(t)\ll 1$ (and corresponding to $1\ll t \ll \tau/4$), 
to  obtain    
\begin{subequations}\elab{FK_smallc}
		\begin{align}
X_1(t)&=4\,e^{-\tau/4}\sinh t\,/c, \enspace  Y_0(t)=2\tan^{-1}(e^{-t}),  
\enspace Y_1(t)=0,
\elab{FK_smallca}
\\ 
\bar{X}_0(t)&=2\tan^{-1}(e^{-\tau/4+t}), \enspace \bar{X}_1(t)=0, 
\enspace \bar{Y}_1(t)=4\,e^{-\tau/4}\cosh(\tau/4-t)\,/c. 
	\end{align}
\end{subequations}
 At this order, the only non-zero contribution to the integral in \eref{Glong} 
comes from the behaviour in Region I. We use \eref{FK_smallca} to obtain that
$| \dot{\bX}(t) - \bu(\bX(t))|^2\sim c^2(\dot{X}_1(t)-X_1(t)\cos Y_0(t) )^2$ and thus 
\beq
\mathscr{G}(c,c_0)\sim \frac{1}{4} \left( 
\frac{32}{\pi}c e^{-\pi/c} -c_0^2 \right)
\eeq
since $c=2\pi/\tau$. 
Solving $\mathscr{G}(\cfk,c_0)=0$ finally gives the approximation
\beq\elab{c1FKapp}
\cfk\sim\frac{\pi}{\mathrm{W}_p\left(32c_0^{-2}\right)}\quad \ \ \textrm{for} \ \ c_0\ll 1.
\eeq
Here, $\mathrm{W}_p$ is the principal branch of the Lambert W function \cite{NIST:DLMF}. 
The above results were previously derived in \cite{TzellaVanneste2014} and included here for completeness.  It is consistent with the bounds of \cite{XinYu2013}. 

We obtain an approximation for $\cg$ in a similar way. The periodic trajectory associated with the variational principle \eref{cgvar} are divided into the same two regions as above. 
The regular expansions are this time more naturally expressed in powers of $c_0$ 
so that in region I where $X(t)\ll 1$, 
we take 
\beq\elab{expansion1_smallc0}
\bX(t)=(0,Y_0(t))+c_0(X_1(t),Y_1(t))+\cdots,
\eeq
 where $X(0)=0$. 
In   region II, $Y(t)\ll 1$ and so we take
\beq\elab{expansion2_smallc0}
\bar{\bm{X}}(t)=(\bar{X}_0(t),0)+c_0(\bar{X}_1(t),\bar{Y}_1(t))+\cdots,
\eeq
where $\bar{X}(\tau/4)=\pi/2$ and once more extend the behaviour over the whole $\tau$ using symmetry. 

\begin{figure*}
\begin{center}
\includegraphics[width=0.47\linewidth]{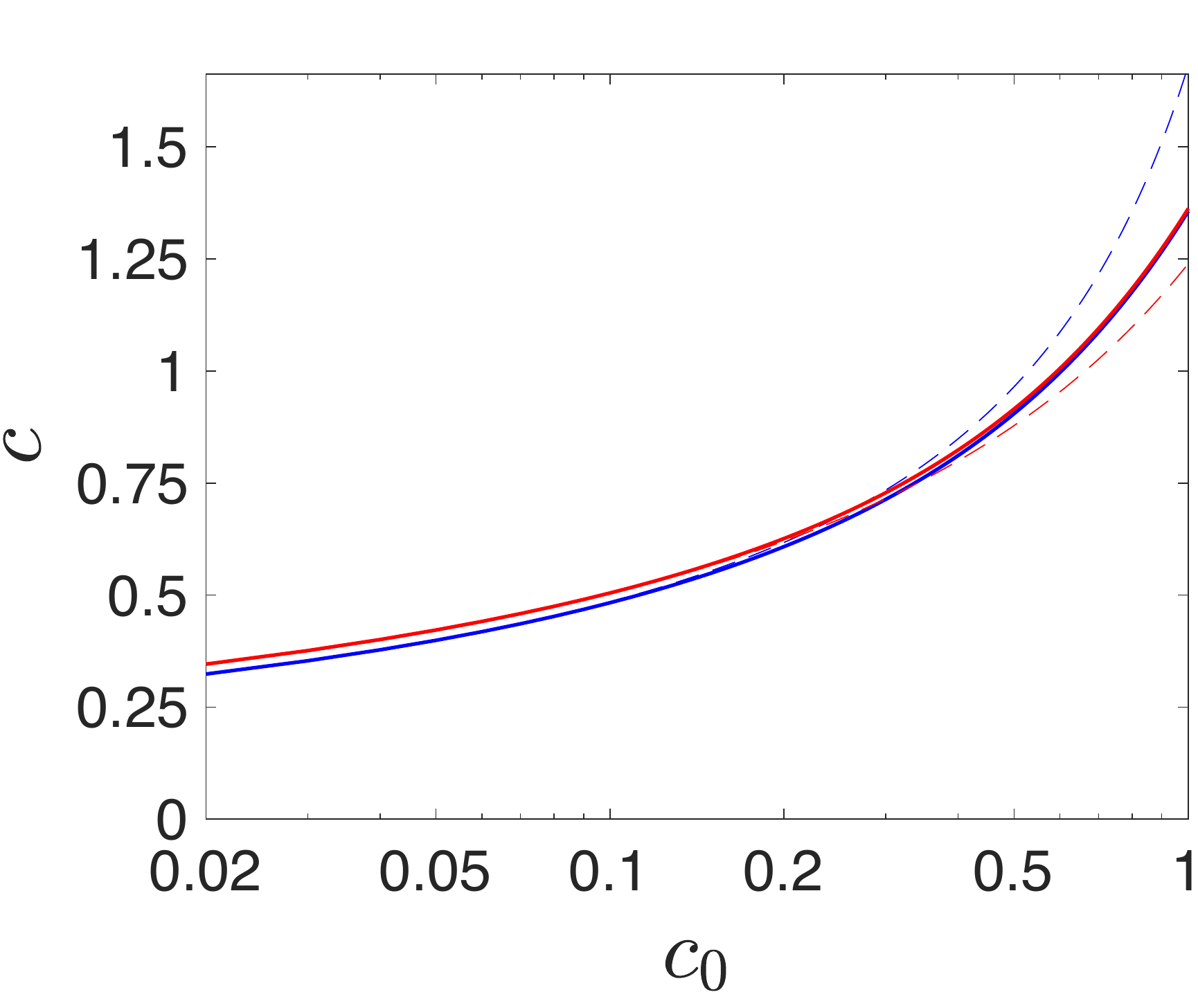}
\includegraphics[width=0.49\linewidth]{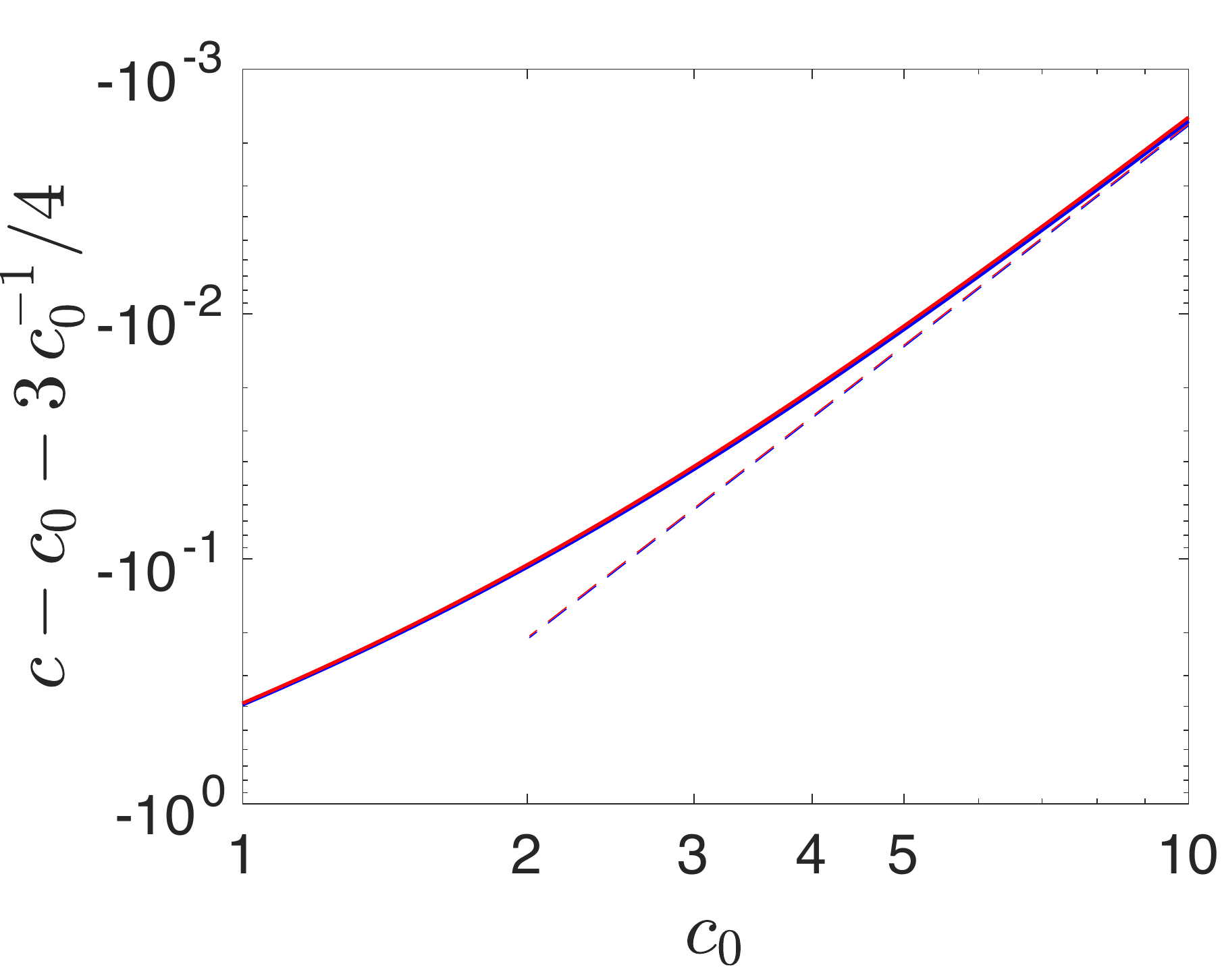}
	\end{center}
\caption{(Color online).  Comparison between numerical  and asymptotic results of the front speed
$c$ associated with  equations \eref{G} (in blue) and  \eref{FK}    (in red).
The numerical results are derived from the   minimisation of \eref{cfkvar} (solid blue  line) and  \eref{cgvar} (solid red  line).
These are juxtaposed against
(left) the small-$c_0$ approximations \eref{c1FKapp}  (dashed blue  line) and \eref{c1Gapp}   (dashed red line) and
(right)  the large-$c_0$ approximations \eref{c2FKapp} (dashed blue  line) and \eref{c2Gapp} (dashed red line).
}
\flab{speed_cell_asymptotics}
\end{figure*}

The periodic trajectory  is now obtained by substituting  \eref{expansion1_smallc0} and \eref{expansion2_smallc0}  inside the pointwise constraint in  \eref{cgvar} from where we obtain equations for each power of $c_0$. This leads to  two   sets of equations
\begin{subequations}
	\begin{align}
	\dot{Y}_0&=-\sin Y_0,\enspace \dot{X}_1=X_1\cos Y_0+\cos\Theta_0,\enspace \dot{Y}_1= -Y_1\cos Y_0 +\sin\Theta_0,\\
 	\dot{\bar X}_0&=\sin \bar{X}_0,\enspace \dot{\bar X}_1=\bar{X}_1\cos \bar{X}_0+\cos\bar{\Theta}_0,\enspace \dot{\bar Y}_1=-\bar{Y}_1\cos \bar{X}_0 +\sin\bar{\Theta}_0,
 \end{align}
 \end{subequations}
where $\Theta_0(t)$ and $\bar{\Theta}_0(t)$  arise when parameterising the constraint \eref{cgvar} in polar coordinates.
The minimum value of $\tau$, denoted by $\tg$, is obtained by maximising $\dot{X}_1(t)$,  $\dot{\bar X}_0(t)$ and $\dot{\bar X}_1(t)$. This gives $\Theta_0(t)=\bar\Theta_0(t)=0$ and leads to 
\begin{subequations}\elab{G_smallc}
		\begin{align}
X_1(t)&=2\cosh t \, \tan^{-1}(\tanh (t/2)), \enspace  Y_0(t)=2\tan^{-1}(e^{-t}), \enspace Y_1(t)=0,\\
\bar{X}_0(t)&=2\tan^{-1}(e^{-\tg/4+t}), \enspace \bar{X}_1(t)=-\tanh(\tg/4-t), \enspace
\bar{Y}_1(t)=\alpha\cosh(\tg/4-t),
	\end{align}
\end{subequations}
since $\cg=2\pi/\tg$, where $\alpha$ is a constant to be determined. 
Matching between the   solutions at $O(c_0)$ in their common region of validity,  given by $X(t),Y(t)\ll 1$  (the same cell corner as above),  yields
an expression for $\cg$.
Using \eref{cgvar}, we deduce that 
\beq\elab{c1Gapp}
\cg=-\frac{\pi}{2 \log \left(\pi c_0/8\right)}\left(1+O(c_0^2)\right),\quad \ \ \textrm{for} \ \ c_0\ll 1
\eeq
and  $\alpha=\pi/2$. The order of the error is estimated  by   matching
the solutions at $O(c_0^2)$ (calculations not shown).
This is qualitatively similar to the expression   obtained in  
 \cite{Abel_etal2002,Cencini_etal2003}  using a heuristic approach and consistent with the rigorous bounds of \cite{XinYu2013}.

Figure \fref{speed_cell_asymptotics} shows that expressions \eref{c1FKapp} and   \eref{c1Gapp}
are in excellent agreement with our numerical solutions; the same is true for expressions \eref{FK_smallc} and \eref{G_smallc} describing the  trajectories (not shown). 
We may use $\mathrm{W}_p(x)=\log(x)-\log\log(x)+o(1)$ as $x\to\infty$ to
further approximate \eref{c1FKapp} as $\cfk\sim-\pi/\left(2 \log (c_0/\sqrt{32}) \right)$.
This approximation highlights the leading-order difference between \eref{c1FKapp} and \eref{c1Gapp}. 
However, this   is only a rough approximation which  
cannot, for instance, capture the non-monotonic behaviour of $\cfk-\cg$  that arises for small  $c_0$ values (not shown).  
Note that both derivations  of \eref{c1FKapp} and   \eref{c1Gapp} 
 tacitly assume that $Y_0(0)=\pi/2$. 
This is easily shown to be the case once the behaviour of the trajectory over the whole (rather than a quarter) spatial period of the flow is taken into account.

\subsubsection{Large-$c_0$ asymptotics} A second asymptotic limit corresponds to $c_0\gg 1$.
We extend  the approach in \cite{TzellaVanneste2014} and take the minimising trajectory  associated with 
the functional in  \eref{Glong}   to be at leading order a straight line  with higher order corrections given by a regular  expansion in $c^{-1}$: 
\beq\elab{expansion1}
\bX(t)=(c t,  Y_0)+c ^{-1}(X_1,Y_1)
+c ^{-2}(X_2,Y_2)
+c ^{-3}(X_3,Y_3)
+c ^{-4}(X_4,Y_4)+\cdots,
\eeq
where $X(0)=0$ and $Y(0)=Y_0$. Here, $Y_0$ is a constant and $X_i(c t)$ and $Y_i(c t)$ are $2\pi$-periodic functions (with zero mean). 
Substituting  \eref{expansion1} into \eref{Glong} gives a
sequence of integrals corresponding to successive powers of $c^{-1}$, obtained using a symbolic algebra package. 
These are in turn minimised 
up to $O(c^{-2})$ with respect to $Y_0$, $X_1(c t)$, $Y_1(c t)$, $X_2(c t)$ and $Y_2(c t)$ (contributions from $X_3(c t)$, $Y_3(c t)$, $X_4(c t)$ 
and $Y_4(c t)$  cancel)
yielding
\beq\elab{FKequation_traj_largec}
Y_0=\pi/2,\enspace
X_1=
Y_2=0,\enspace
Y_1=-2\sin(c t), \enspace X_2=-\frac{3}{8}\sin(2c t).
\eeq
Introducing \eref{FKequation_traj_largec} into \eref{Glong}    we obtain 
\beq
\mathscr{G}(c,c_0 )=\frac{1}{4}\left( c ^2-\frac{3}{2} +\frac{87}{32}c ^{-2} - c_0^2 \right) +O(c ^{-4}),
\eeq
after a few manipulations. 
This leads to the asymptotics of the speed
\beq\elab{c2FKapp} 
\cfk  
= 
c_0
\left(1+\frac{3}{4}c_0^{-2}-\frac{105}{64}c_0^{-4}+O(c_0^{-6})\right)
\quad \ \ \textrm{for} \ \ c_0\gg 1, 
\eeq
with the first two terms previously derived in \cite{TzellaVanneste2014}.

In a similar manner,  the minimising trajectory associated with 
the variational principle  \eref{cgvar} for \eref{G} is at leading order   a straight line.  
Using the alternative variational characterisation \eref{cgvar2}, we write the trajectory in terms of $x$ 
and take a  regular expansion in powers of $c_0^{-1}$: 
\begin{subequations}\elab{expansion3}
	\begin{align}
     T(x)&=c_0^{-1}(x+c_0^{-1}{T}_1+c_0^{-2}{T}_2
	 +c_0^{-3}{T}_3+c_0^{-4}{T}_4)+\cdots,\\
	 Y(x)&={Y}_0+c_0^{-1}{Y}_1+c_0^{-2}{Y}_2
	 +c_0^{-3}{Y}_3+c_0^{-4}{Y}_4+
	 \cdots,
 	\end{align}
\end{subequations}
where $Y(0)={Y}_0$. The  ${Y_i}$'s are $2\pi$-periodic functions  satisfying ${Y}_i(0)=0$ while ${T}_i(0)=0$ for all $i\geq 1$. 
We  substitute these inside the pointwise constraint in  \eref{cgvar2} 
from where we obtain equations for each power of $c_0^{-1}$.
This leads to  expressions for ${T}_i'(x)$ which are  in turn used to minimise $\int_0^{2\pi}{T}_i'\,\d x$. 
Up to $O(c_0^{-2})$ and after a few manipulations 
{carried out with a symbolic algebra package} we obtain that
\begin{subequations}\elab{Gequation_traj_largec}
\elab{expansion2}
	\begin{align}
     {T}_1&={T}_3=0,\enspace
	 {T}_2=-3x/4+f(x), \enspace
	 {T}_4(x)=145x/64+g(x),
	 \\
	 {Y}_0&=\pi/2,\enspace
	 {Y}_1=-2\sin x,\enspace
	  {Y}_2=0, 
	\end{align}
\end{subequations}
where $f(x)=5\sin(2x)/8$ and $g(x)=-Y_3(x)\cos x-143\sin(2x)/96+17\sin(4x)/768$ are $2\pi$-periodic and therefore do not contribute to the value of $\tg$.
Note that the difference between the two trajectories obtained in \eref{FKequation_traj_largec}
and \eref{Gequation_traj_largec} only appears at $O(c_0^{-2})$. 
We finally  
use \eref{cgvar2} to deduce that
\beq\elab{c2Gapp} 
\cg  
= 
c_0
\left(1+\frac{3}{4}c_0^{-2}-\frac{109}{64}c_0^{-4}+O(c_0^{-6})\right)
\quad \ \ \textrm{for} \ \ c_0\gg 1. 
\eeq

Comparing expressions \eref{c2FKapp} and  \eref{c2Gapp}  confirms that   
the difference between the front speeds for the \eref{FK} and \eref{G} equation is very small: equation \eref{G} only slightly underpredicts the front speed.
This is confirmed in Figure \fref{speed_cell_asymptotics} which focuses on verifying  \eref{c2FKapp} and  \eref{c2Gapp}. 
It is clear that the two approximations \eref{c2FKapp} and \eref{c2Gapp}  are in excellent agreement with the numerical results; however, 
they are too close apart to distinguish.

\subsection{Perturbed cellular flow}\slab{perturbedcellular}

\begin{figure*}
		\begin{center}
			\begin{minipage}{\linewidth}
				\begin{minipage}{0.49\linewidth}
					\begin{center}
	 \includegraphics[width=\linewidth]{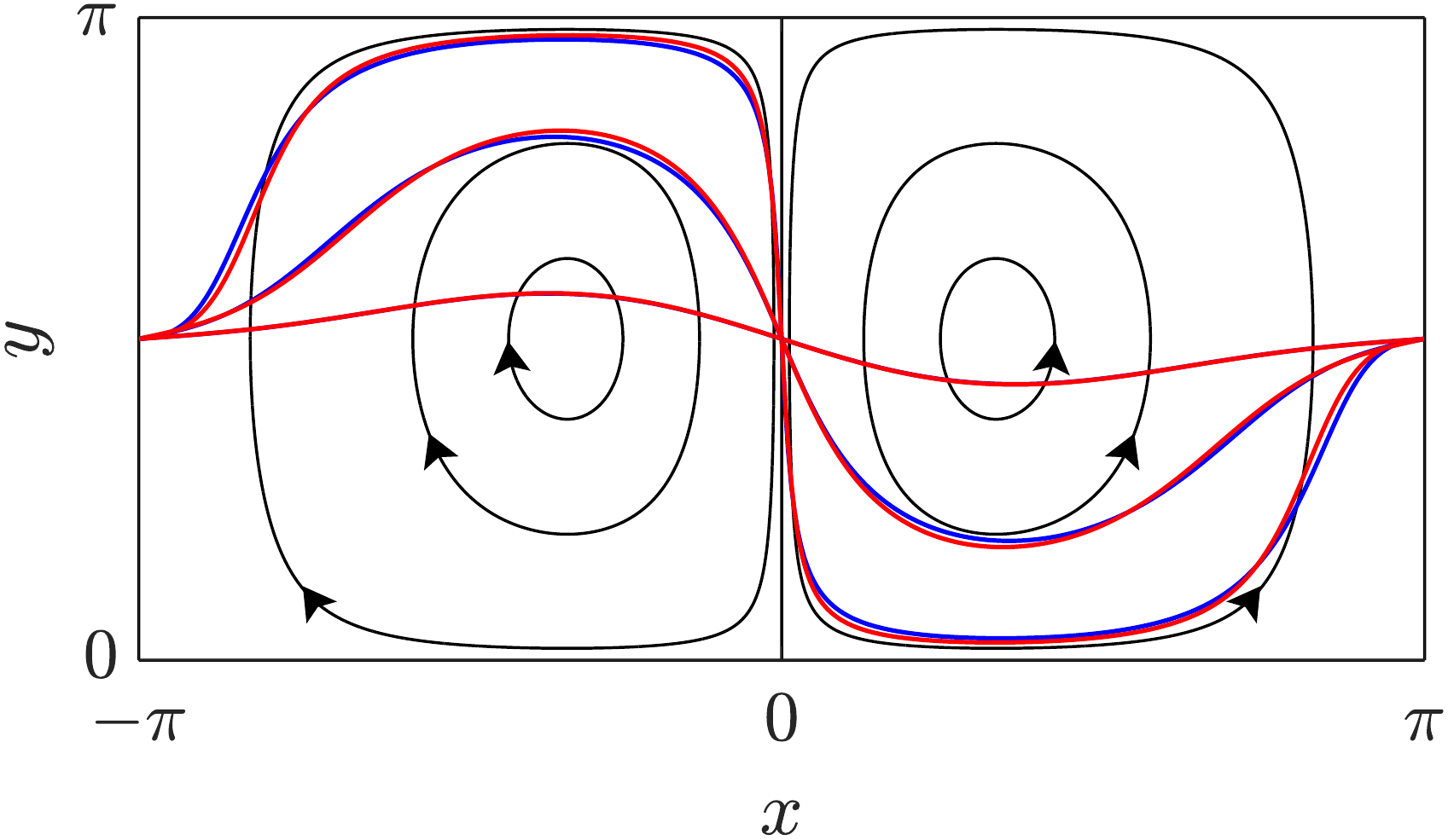} 
	 \end{center}
	  \end{minipage}
	  \begin{minipage}{0.49\linewidth}
		 \begin{center}
	 \includegraphics[width=\linewidth]{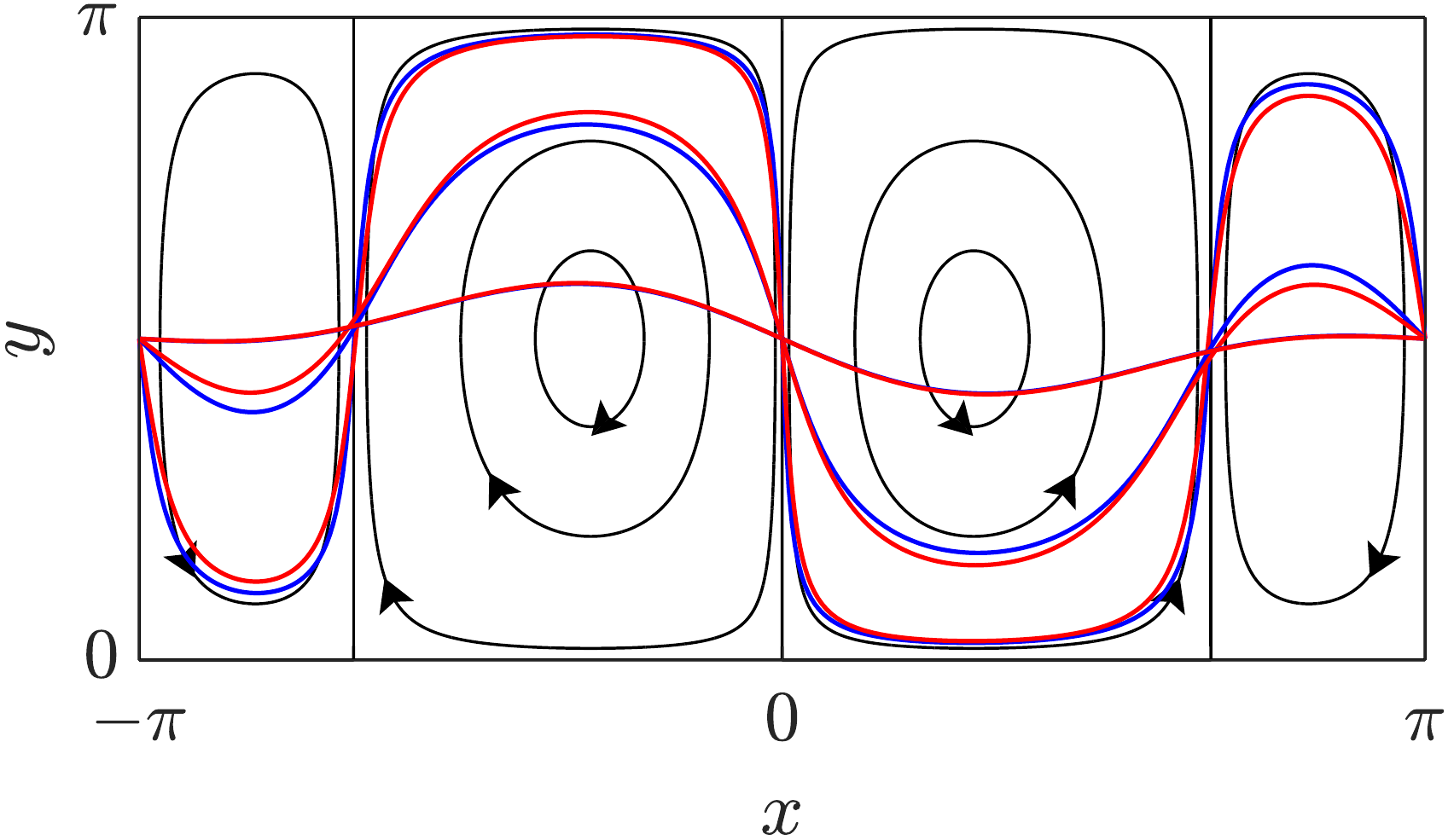}
	 \end{center}
	 \end{minipage}
	\\[5 pt]
	\begin{minipage}{0.49\linewidth}
	\begin{center}
	 \centerline{(a) $A=0.5$, $U=0$}
	 \end{center}
	  \end{minipage}
	  \begin{minipage}{0.49\linewidth}
		 \begin{center}
	  \centerline{(b) $A=1$, $U=0$}
	 \end{center}
	 \end{minipage}
	 \vfill
	 \begin{minipage}{0.49\linewidth}
	 \includegraphics[width=\linewidth]{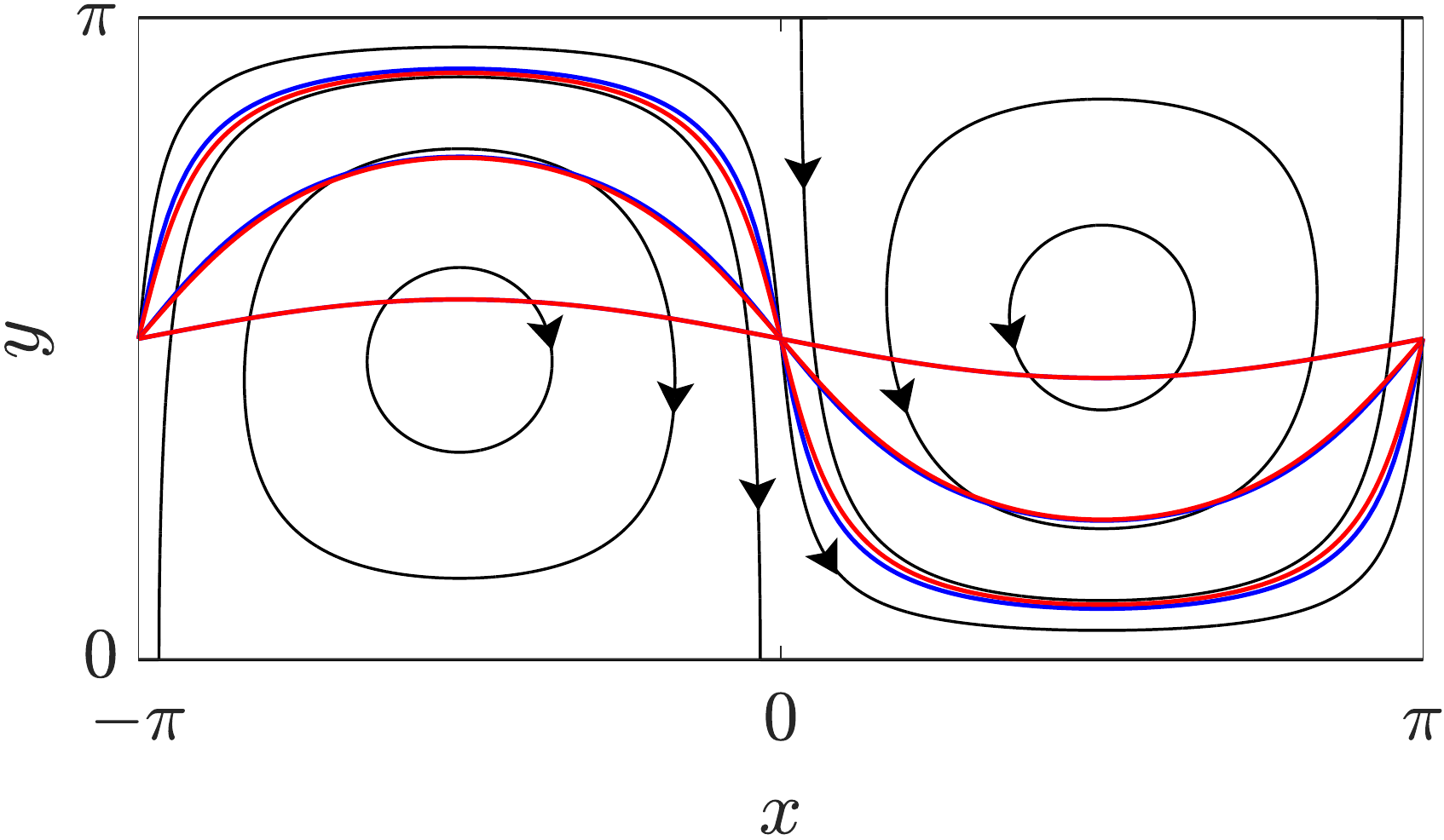} 
	  \end{minipage}
	   \begin{minipage}{0.49\linewidth}
	 \includegraphics[width=\linewidth]{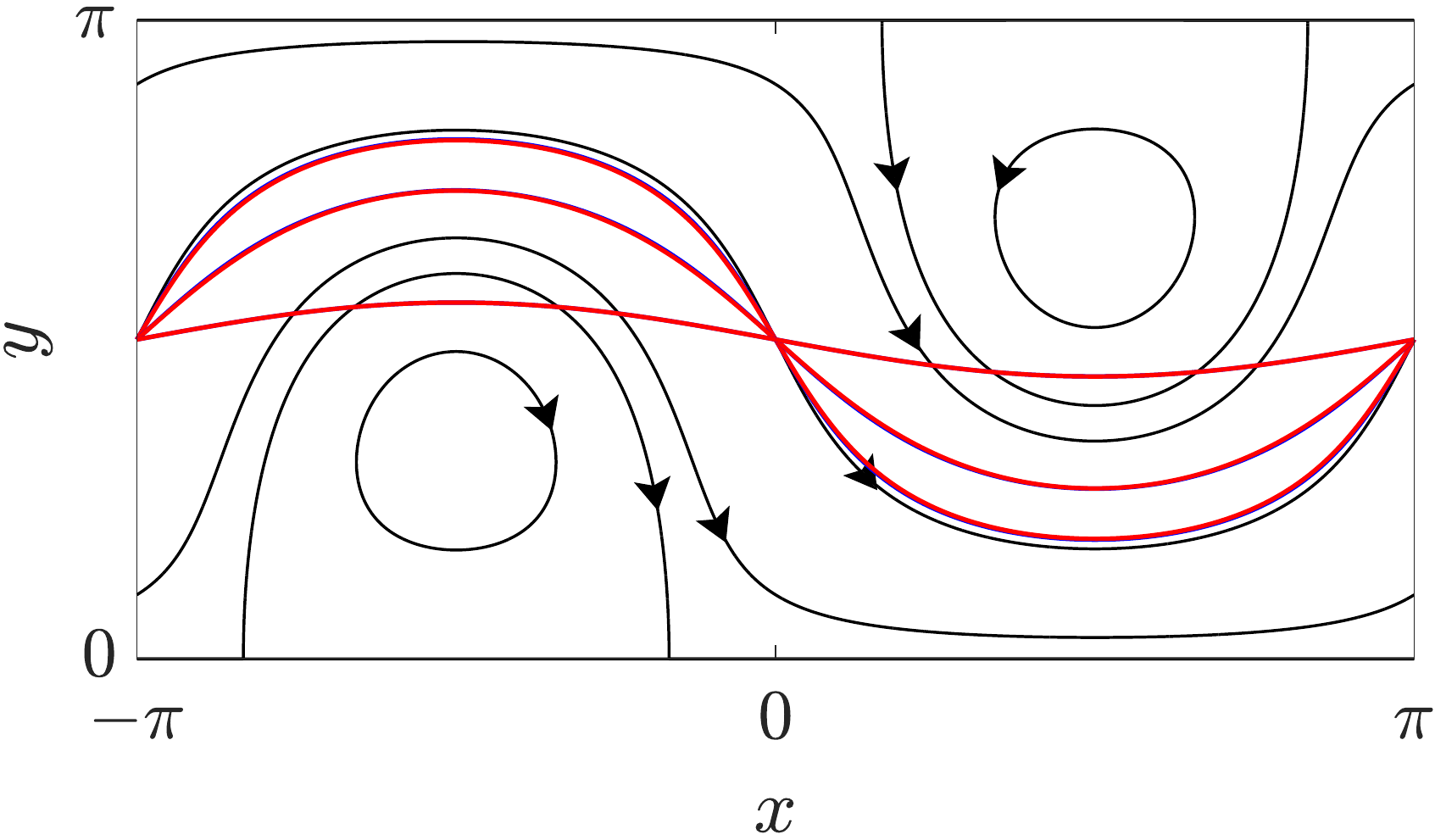}
	 \end{minipage}
 	\\[5 pt]
 	\begin{minipage}{0.49\linewidth}
 	\begin{center}
 	 \centerline{(c) $A=0$, $U=0.1$}
 	 \end{center}
 	  \end{minipage}
 	  \begin{minipage}{0.49\linewidth}
 		 \begin{center}
 	  \centerline{(d) $A=0$, $U=0.5$}
 	 \end{center}
 	 \end{minipage}
 	 \begin{minipage}{0.49\linewidth}
 	 \includegraphics[width=\linewidth]{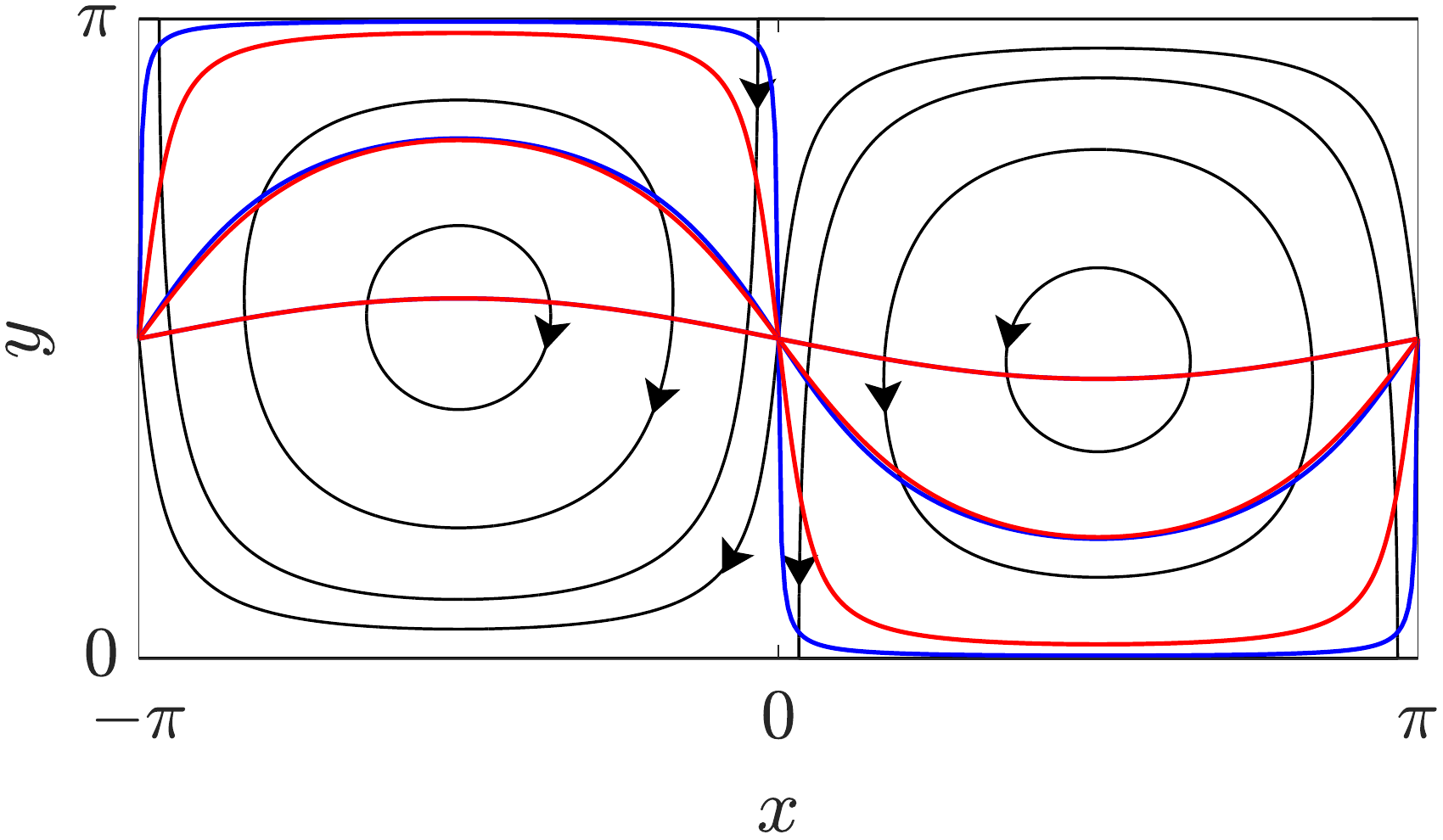} 
 	  \end{minipage}
 	   \begin{minipage}{0.49\linewidth}
 	 \includegraphics[width=\linewidth]{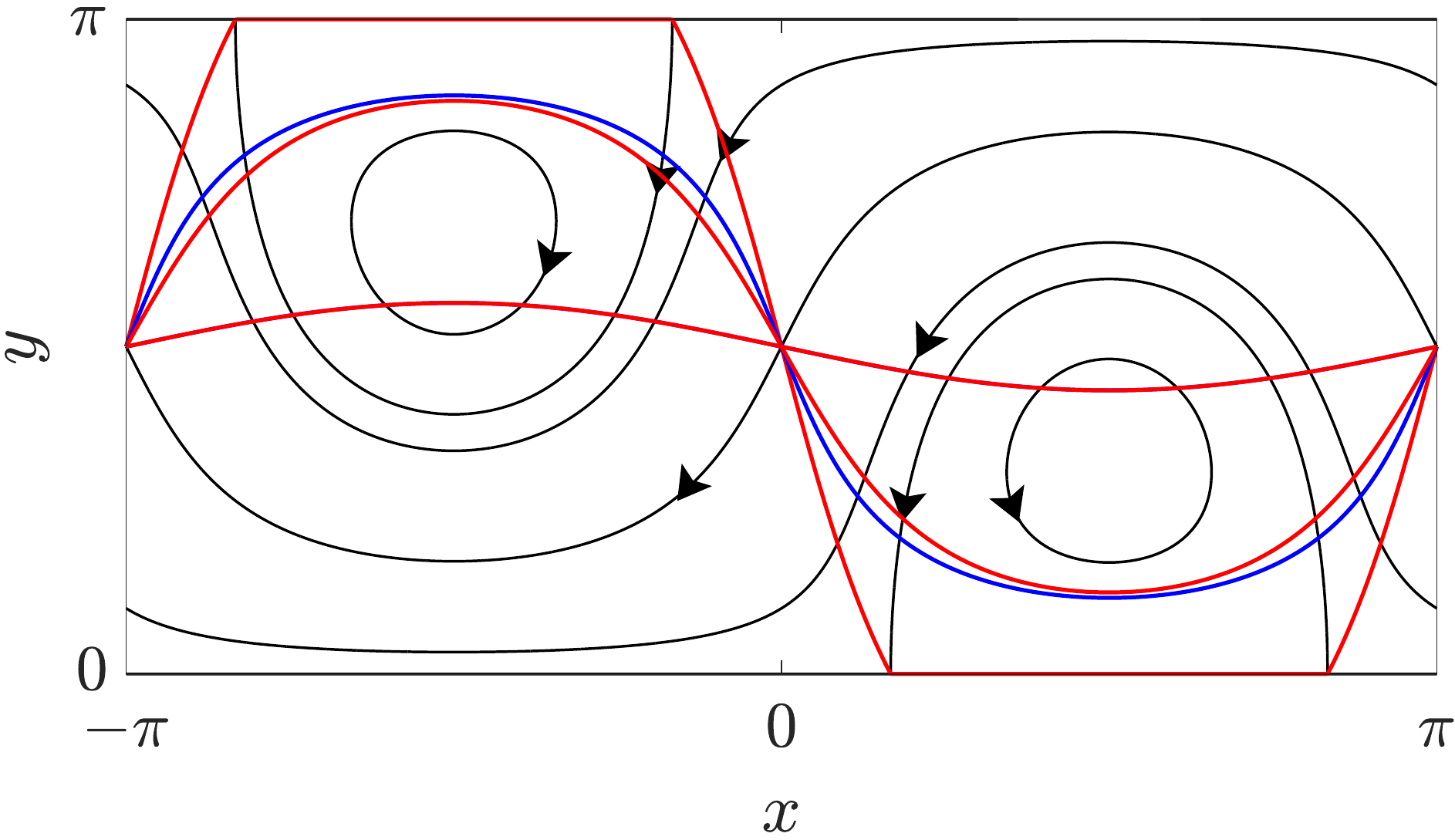}
 	 \end{minipage}
  	\\[5 pt]
  	\begin{minipage}{0.49\linewidth}
  	\begin{center}
  	 \centerline{(e) $A=0$, $U=-0.1$}
  	 \end{center}
  	  \end{minipage}
  	  \begin{minipage}{0.49\linewidth}
  		 \begin{center}
  	  \centerline{(f) $A=0$, $U=-0.5$}
  	 \end{center}
  	 \end{minipage}
	 	 \end{minipage}
		\end{center}
\caption{(Color online).  Streamlines (thin  black lines) of  the closed cellular flow with streamfunction \eref{streamfunction} 
with $A \not=0$ and $U = 0$ (top row) and with $A=0$ and $U \not=0$ (middle and bottom rows),  
and corresponding periodic trajectories for \eref{FK} (minimising \eref{cfkvar}, thick blue  lines) and   \eref{G} (minimising \eref{cgvar}, thick red  lines). For the top and middle rows, the minimising trajectories are plotted for $c_0=0.1,\, 1$ and $10$ (cf.\ Figure \fref{trajectories_cell} for $A=U=0$). For panel (e), with $U=-0.1$, there is no right-propagating \eref{G} front for $c_0=0.1$  and the three values $c_0=0.11,\, 1$  and $10$ have been used. For panel (f), with $U=-0.5$, there are no right-propagating \eref{FK} and \eref{G} fronts for 
$c_0=0.01$ and the values $c_0=0.19,\, 1$  and $10$ have been used; there is no right-propagating \eref{G} front for $c_0=0.19$. Note that the \eref{FK} and \eref{G} trajectories are often indistinguishable for the larger values of $c_0$.}
\flab{trajectories_small_scale}
\end{figure*}

We now investigate the effect of perturbing the basic cellular flow by taking for $A\neq 0$ 
in the streamfunction \eref{streamfunction}, keeping $U=0$. 
The perturbation breaks a  symmetry of the streamfunction.
Characteristic examples of trajectories associated with \eref{FK} and \eref{G} are shown in Figure \fref{trajectories_small_scale} (top row) 
for two values of $A$ corresponding to distinctly different flow topologies.
The trajectories remain symmetric for the transformation $(x,y)\mapsto (-x,\pi-y)$.
Qualitatively, they are similar to those obtained for $A=0$, following closely the straight line $y=\pi/2$ when $c_0$ is large and the separatrix when $c_0$ is small. Despite the more complex flow structure, the difference between the \eref{FK} and \eref{G}  trajectories remains small.

Figure \fref{speed_small_scale_open} (top) shows the behaviour of 
$\cfk$  as a function of $c_0$.   
For $0<A\leq 1$, the value of $\cfk$ does not greatly differ from the 
corresponding value 
 obtained for $A=0$. 
A significant difference is obtained for $A=5$.
For large $c_0$, $\cfk$ increases 
quadratically with $A$. This can be shown by generalising the asymptotic result \eref{c2FKapp} to find, after a lengthy computation, that    
 $\cfk=c_0(1+(12+9A^2)c_0^{-2}/16-3(280 + 504 A^2 + 101 A^4)c_0^{-4}/512+\cdots)$ for $c_0\gg 1$ and $A=o(c_0)$.
Expansions 
 \eref{expansion1_smallc0a} and \eref{expansion2_smallc0a} 
 can in principle also be generalised to provide an explicit expression for $\cfk$
when $c_0\ll 1$. However, the computation becomes very involved, especially for $A \ge 1/2$ when the number of hyperbolic stagnation points is doubled; we have not attempted this computation. 

\begin{figure*}
		\begin{center}
		\includegraphics[width=0.48\linewidth]{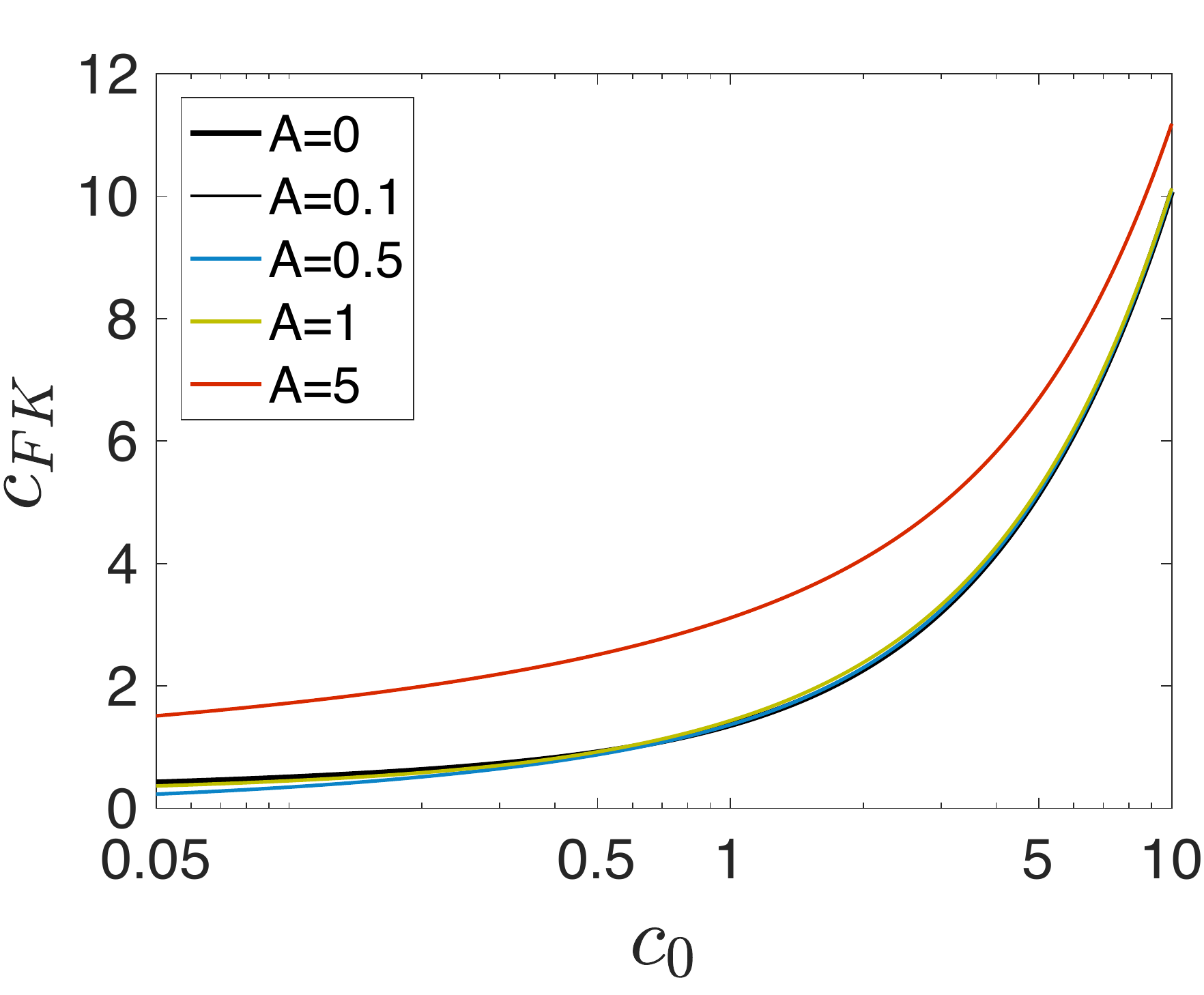}
			\end{center}
		\vfill
			\begin{center}
			\begin{overpic}[width=0.49\linewidth]{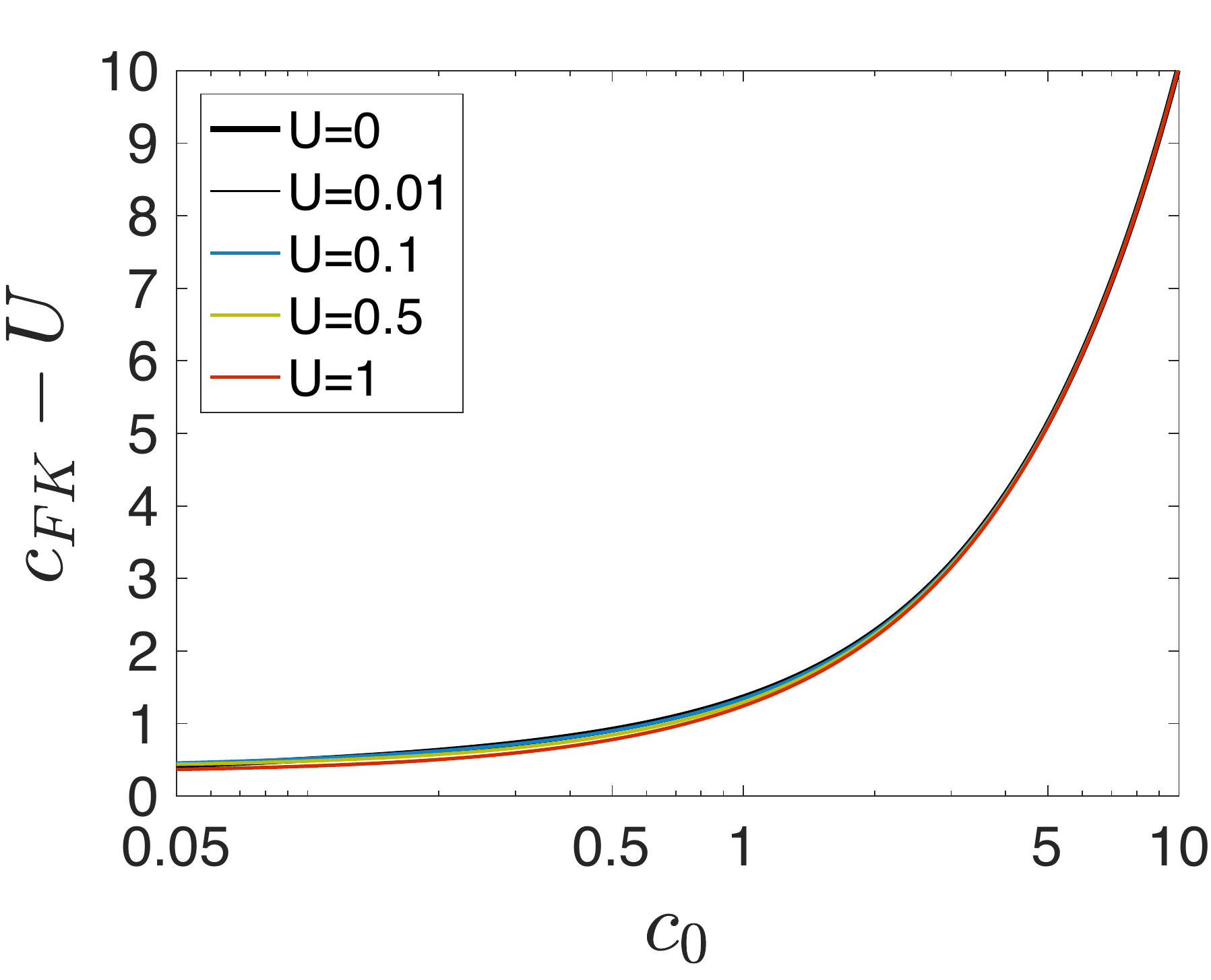}
				\put(40,40){\includegraphics[scale=0.155]{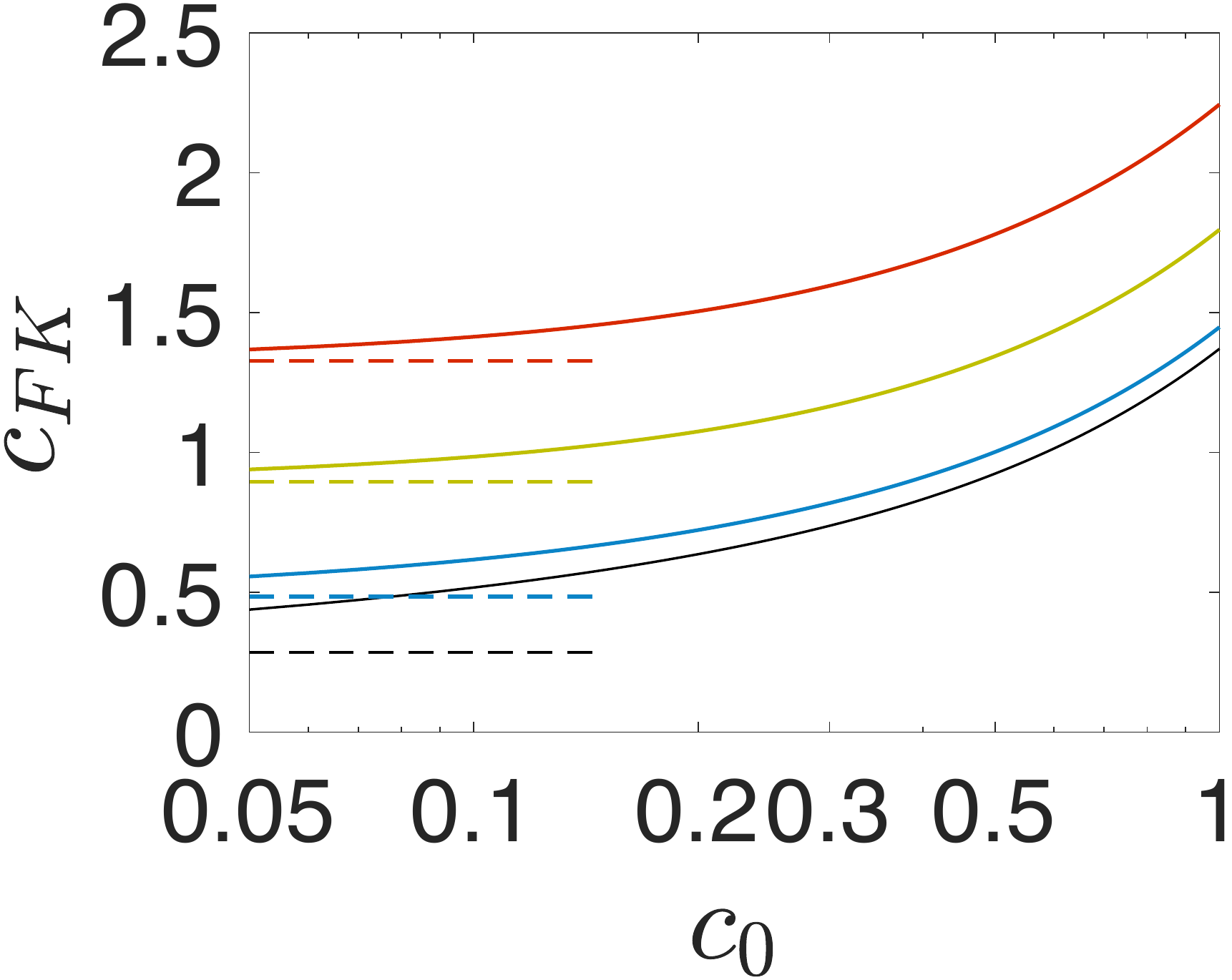}}
			\end{overpic}	
			\begin{overpic}[width=0.49\linewidth]{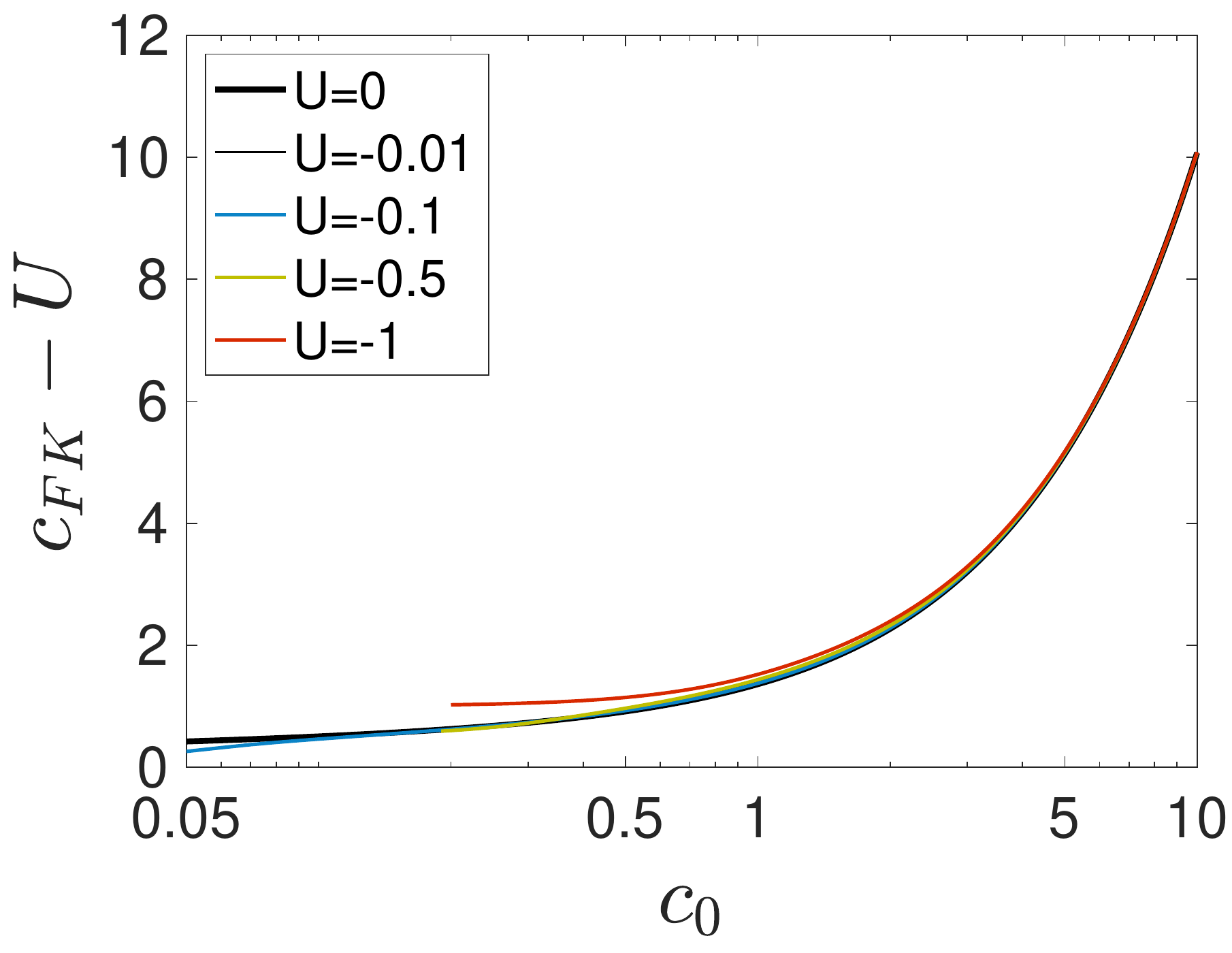}
				\put(41,38){\includegraphics[scale=0.153]{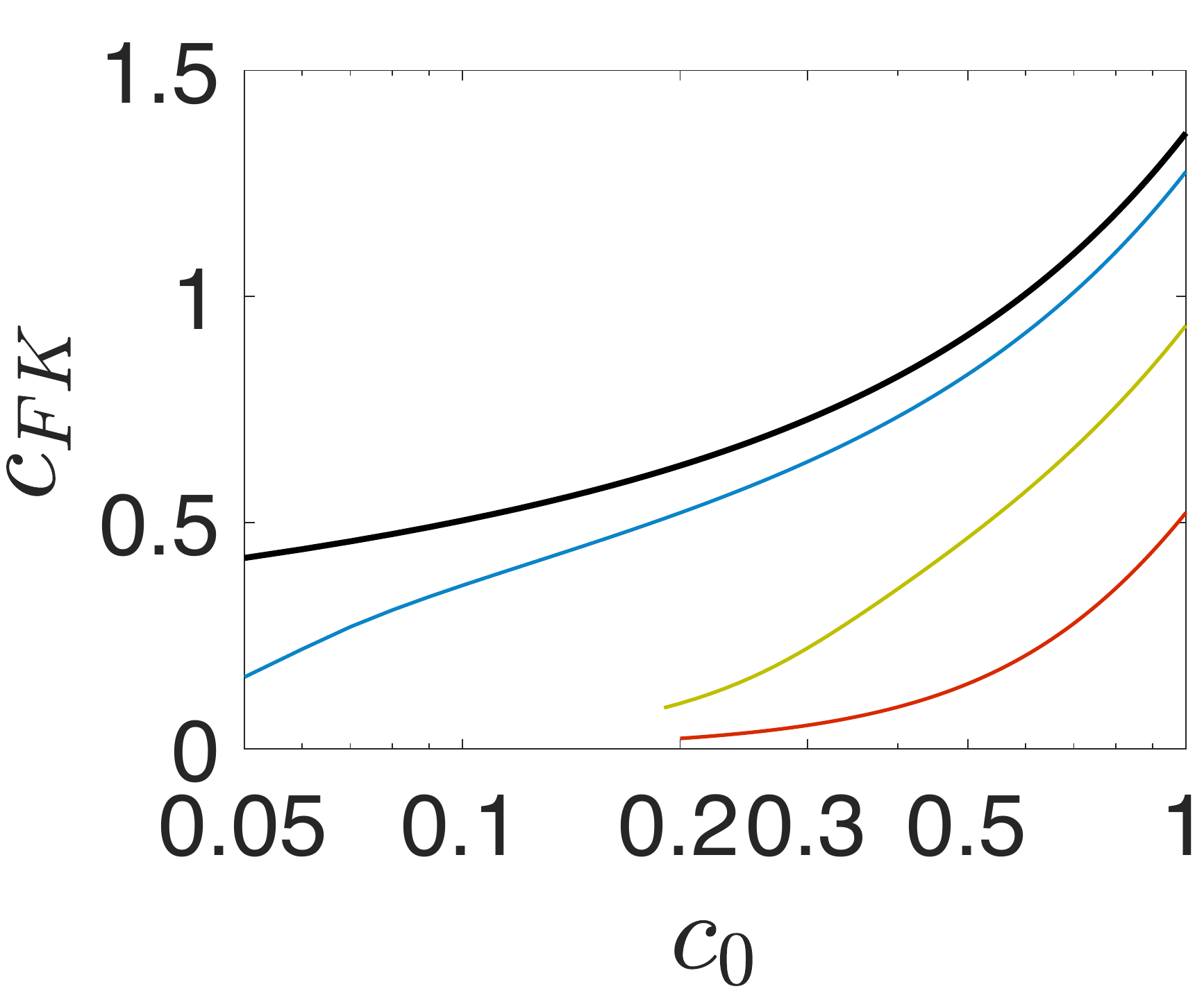}}
			\end{overpic}
			\end{center}
\caption{(Color online).
Front speed $\cfk$ 
 associated with equation \eref{FK} 
plotted as a function of 
the bare speed $c_0$ for the flow with streamfunction \eref{streamfunction}
for (top row) various values of $A$ with $U=0$ and (bottom row)
for various values of $U$ with $A=0$ 
($\cfk$ is shifted by $U$).
The insets focus on the small-$c_0$ behaviour of $\cfk$ (solid lines)
and (left) how this compares with $c_{\+}(U)$ obtained from \eref{cplus} (dashed lines).
 }
\flab{speed_small_scale_open}
\end{figure*}

Figure \fref{speed_small_scale_open_flow} (top left) shows the   difference  
between the two front speeds $\cfk$ and $\cg$ 
as a function of $c_0$ and 
for a number of values of $A$.  
This varies non-monotonically with $c_0$, with a peak 
whose location is not simply related to  $A$. 
 We  observe that for  values of $c_0$ as large as $1$, there is no clear relation between this   difference and the value of $A$. 
 For larger values of $c_0$, the  difference increases with $A$. 
 This can be shown 
 using the generalisations of  the asymptotic approximations \eref{c2FKapp} and \eref{c2Gapp} which give 
 $(\cfk-\cg)/f(A)=(1+O(c_0^{-1}))c_0^{-3}$,
 where $f(A)=(16+538 A^2+ A^4)/256$,  for  $c_0\gg 1$ and $A=o(c_0)$. 
The relative difference between the two front speeds is shown in Figure \fref{speed_small_scale_open_flow} (top right). 
For the values of $c_0$ considered here, the maximum relative difference between 
$\cfk$ and $\cg$ corresponds to $9\%$,  achieved for $A=1/2$
and $c_0=0.05$. This is not significantly different to 
the maximum relative difference of $5.5\%$
obtained for $A=0$. 

\begin{figure*}
	\begin{center}
	\includegraphics[width=0.45\linewidth]{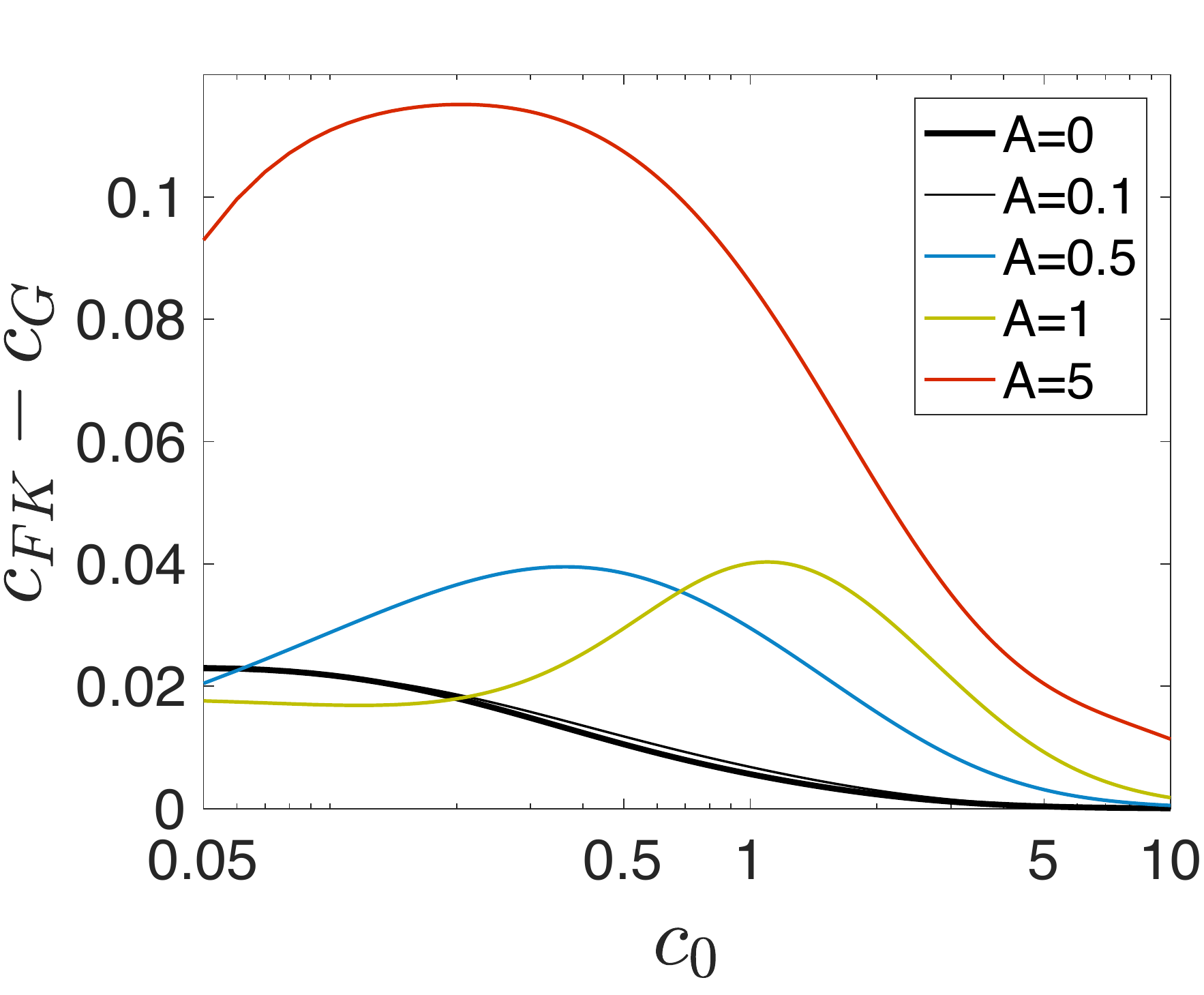}
	\includegraphics[width=0.46\linewidth]{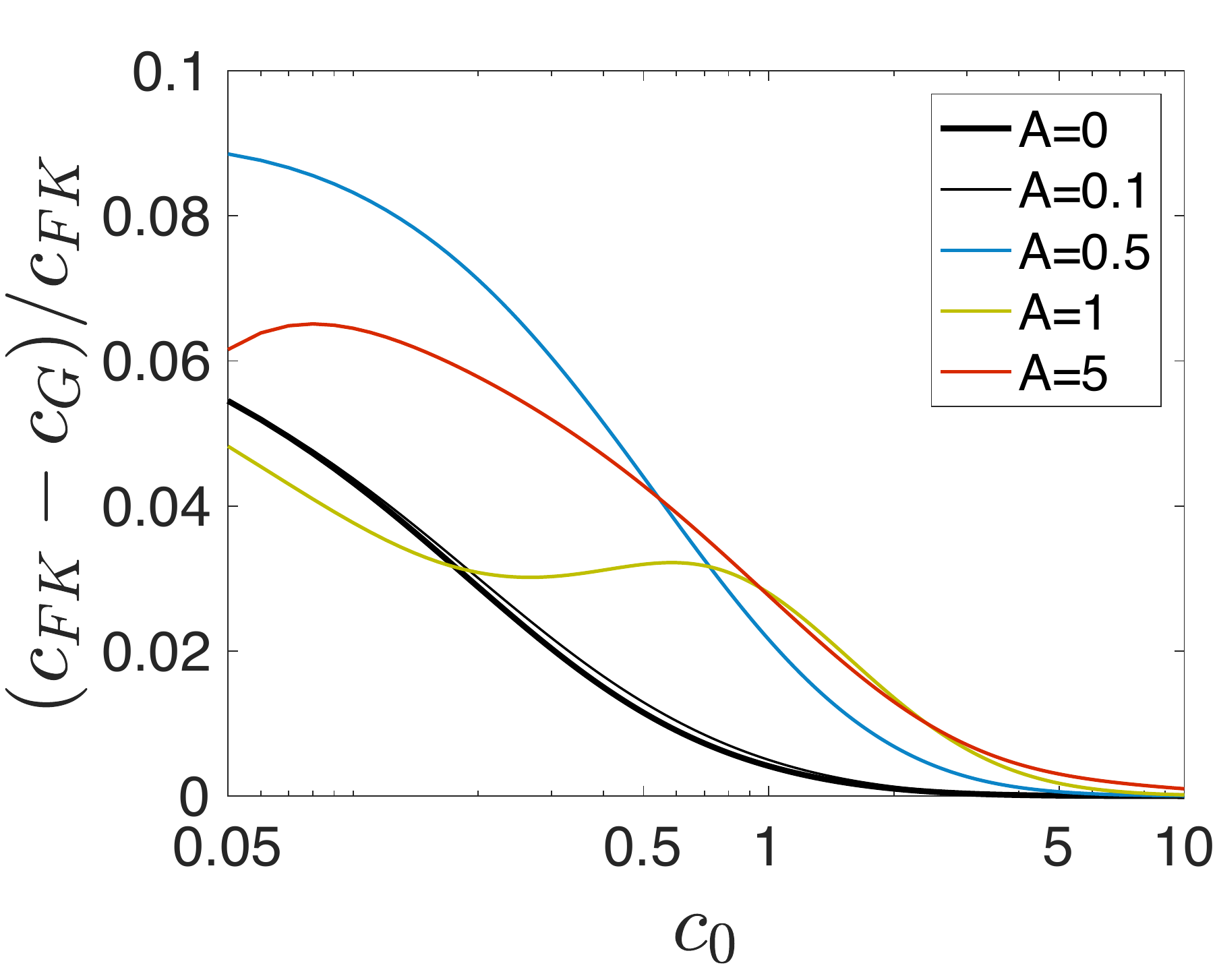}
	\includegraphics[width=0.45\linewidth]{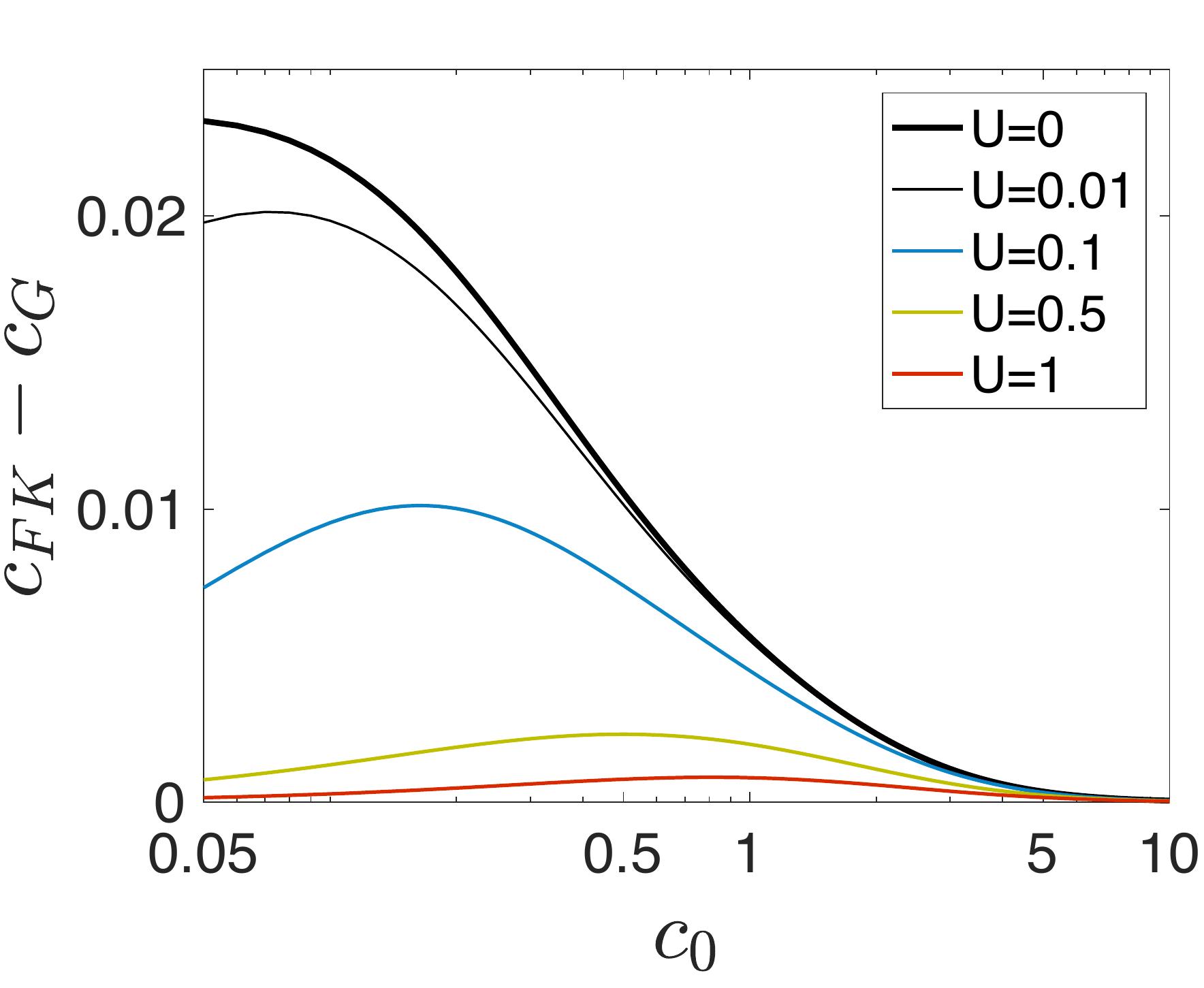}
	\includegraphics[width=0.46\linewidth]{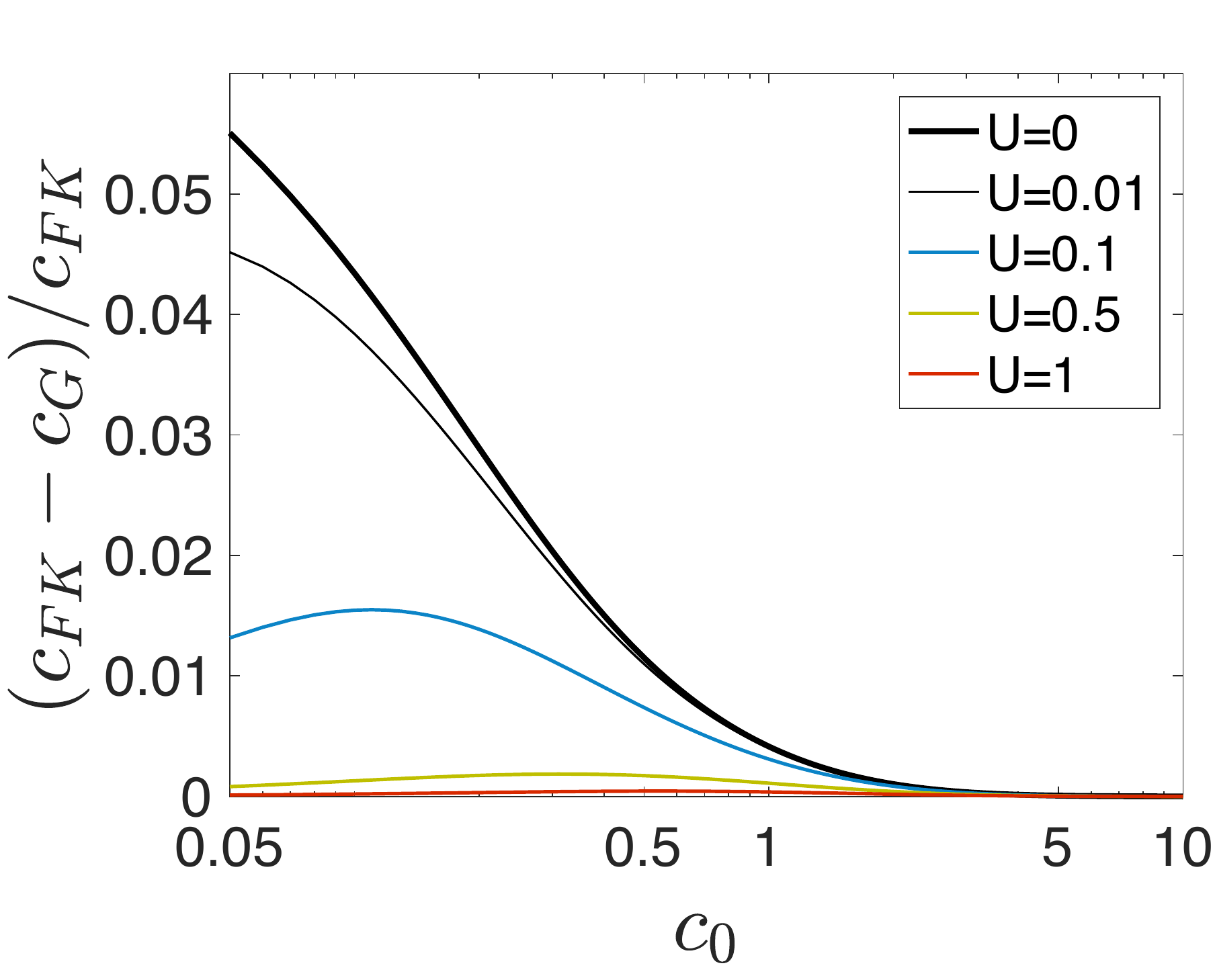}
	\includegraphics[width=0.45\linewidth]{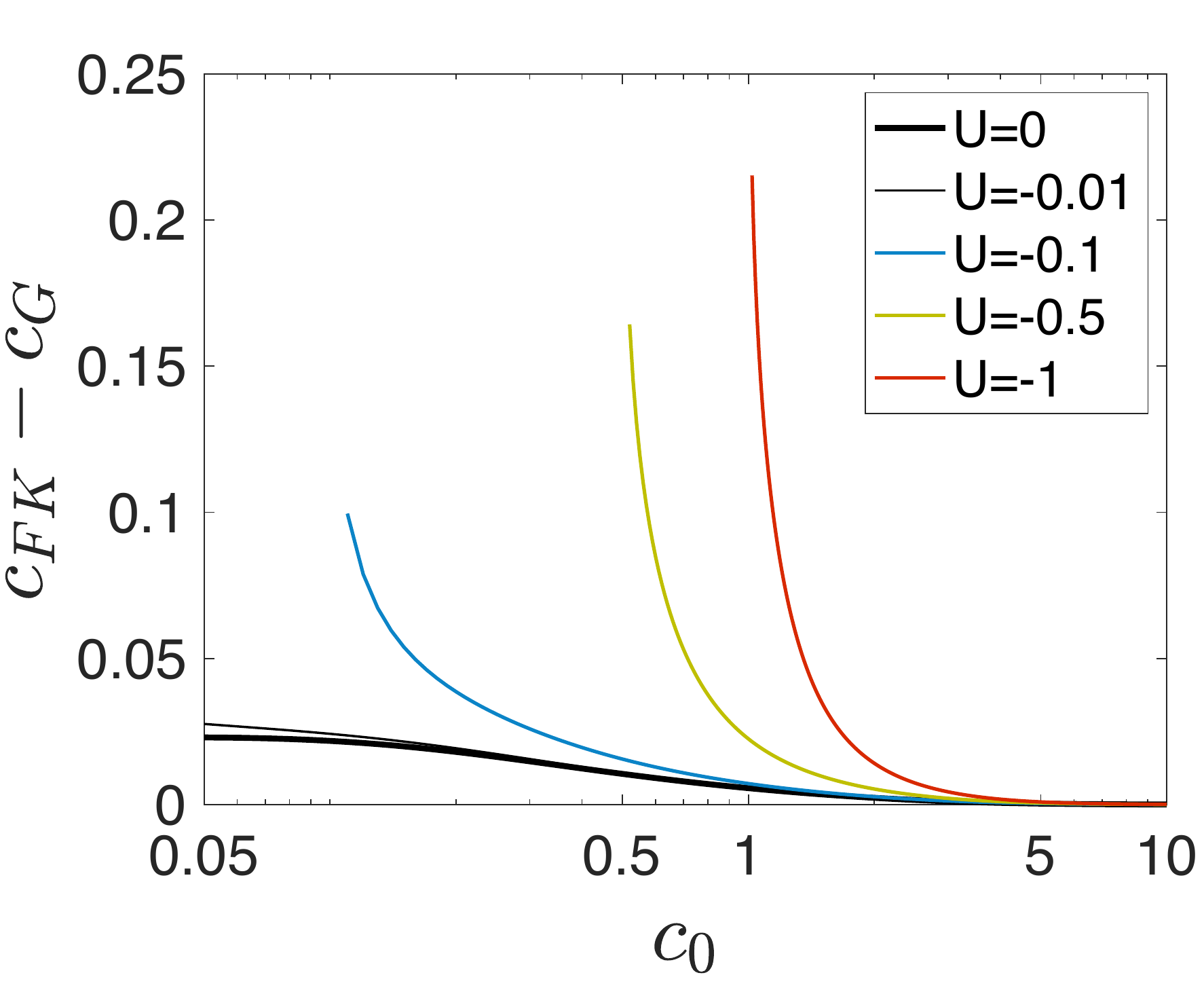}
	\includegraphics[width=0.46\linewidth]{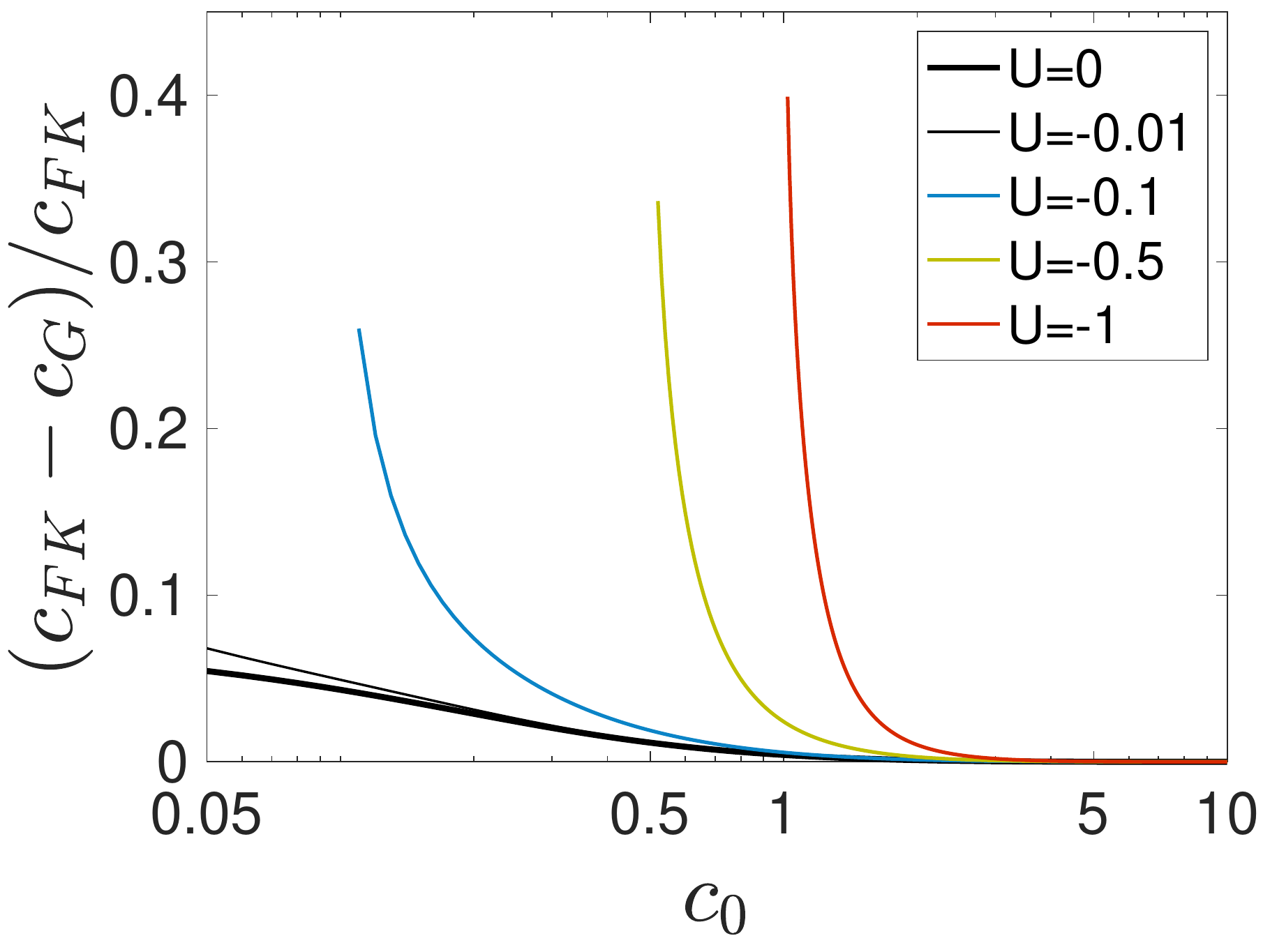}
	\end{center}
\caption{(Color online).  
Effect of the flow with streamfunction \eref{streamfunction} on the (left column) difference and (right column) relative difference between the front speed $\cfk$ associated with equation \eref{FK}
 and the front speed $\cg$ associated with equation \eref{G}. 
 These are plotted
as a function of the bare speed $c_0$ for (top row) various values of $A$ with $U=0$ and (middle and bottom rows) various values of $U$ with $A=0$.  
The values of $\cfk$ and $\cg$ are respectively
derived from the numerical minimisation of the variational principles \eref{cfkvar} and \eref{cgvar}. 
As $c_0\to-U>0$, $\cg\to 0^+$ so that the relative difference tends to $1^-$ (bottom right).
 }
\flab{speed_small_scale_open_flow}
\end{figure*}

\subsection{Effect of a mean flow} \label{sec:meanflow} 
The behaviour of the solutions is strongly affected by the presence of a
constant mean flow, when 
the flow contains a mixture of open and closed streamlines.  
We explore this by computing minimising trajectories and front speeds for $U \neq 0$ and $A=0$.
Figure \fref{trajectories_small_scale} shows characteristic examples of the minimising  trajectories
obtained for  different values of $U>0$ (middle row)  and $U<0$ (bottom row). 
These trajectories are clearly  invariant under the transformation $(x,y) \mapsto (-x,\pi-y)$. 

For small values of $c_0$ and $U=O(1)>0$, the minimising  trajectories  closely follow the
 open streamline  with the maximum average horizontal speed 
 $c_{\+}(U)$, say, situated in the 
 middle of the channel, which suggests that $\cfk \sim c_{\+}$.  It can be shown that
 \beq\elab{cplus}
c_{\+}(U) 
= \frac{2\pi}{\tau_{\+}(U)},\ \ 
\text{where} \ \ 
\tau_{\+}(U)=4 \int_0^{z_{\+}} \frac{d z}{\left(\cos^2 z - U^2 z^2\right)^{1/2}}
 \eeq
and $0 \le z_{\+} \le \pi/2$ is the solution of $\cos z_{\+} = U z_{\+}$. 
A comparison between $\cfk$ and $c_{\+}$ in Figure \fref{speed_small_scale_open} (bottom left, inset) confirms the validity of this prediction, although convergence as $c_0 \to 0$ is slow. The prediction is not applicable when $U=O(c_0)$, however. This is because the travel time along the fastest open streamline increases (like $4 \log (1/U)$) and trajectories entering the closed streamlines (analogous to the the optimal trajectories obtained for $U=0$ as $c_0 \to 0$) become more favourable.

For large values of $c_0$, we can extend  
the  asymptotic expansion \eref{expansion1} to account for $U>0$ to deduce that, at leading order, $\cfk$ is simply shifted by $U$ compared with its value when $U=0$. 
Figure \fref{speed_small_scale_open} (bottom  left) confirms 
this behaviour by showing 
$\cfk-U$ as a function of $c_0$  for different values of $U$ (including $U=0$) and exhibiting the expected collapse of curves for large $c_0$.

Figure \fref{speed_small_scale_open_flow} (middle row) 
compares the two front speeds $\cfk$ and $\cg$ for $U>0$. The difference in speed decreases as $U$ increases and is maximum for an intermediate value of $c_0$ for $U \not=0$ as well as for $U=0$. 
The relative difference between the two front speeds is very small:
for the values of $c_0$ considered here, the maximum relative difference between 
$\cfk$ and $\cg$ is approximately $4.5\%$,  achieved for $U=0.01$
and $c_0=0.05$. 
For $U\gtrsim 0.2$, the maximum relative difference is for all values of $c_0$ less than $1\%$.
When $U>1$, the flow is entirely composed of open streamlines and therefore
 similar to a shear flow. As a result the two front speeds are nearly identical.

For  $U<0$ (bottom row of Figure \fref{trajectories_small_scale}), the mean flow opposes the right propagation of the front, and the minimising trajectories avoid regions of strong flow. For small values of $c_0$, 
they follow closely the cell boundary and differ markedly between 
the   \eref{FK} and \eref{G} cases. For sufficiently small $c_0$, the fronts cease to propagate to the right. For  \eref{G}, 
\eref{stat_G} indicates that there is no right-propagating front for $c_0 \le  -U - \min_{x}\max_y\sin x \cos y = -U$.
 Our numerical results suggest that right-propagating fronts do exist for all $c_0>-U$. 
Figure \fref{trajectories_small_scale} (bottom, left) shows 
the behaviour of the  minimising trajectory 
associated with equation \eref{G} 
obtained
near the stationary \eref{G} front limit 
for  $U=-0.1$ and $c_0=0.11$. 
This is  characterised by near-vertical segments at $x=0,\pm \pi$ and $y=\pi/2$ where $\bs{u}=(-U,0)$ and the pointwise constraint  in  \eref{cgvar} imposes that $\dot x$ be small. 
For \eref{FK}, right-propagating fronts are obtained for values of $c_0$ smaller than $-U$. %
For instance, for $U=-0.5$, we find  a nearly stationary front, with very small (positive) $\cfk$, for $c_0=0.19$. The corresponding minimising trajectory is 
shown in Figure \fref{trajectories_small_scale}.

A more complete description is provided by Figure  \fref{speed_small_scale_open} (bottom right) which
shows $\cfk$ for a wide range of values of $c_0$, reaching close to  stationary \eref{FK} fronts as $\cfk \to 0$ (inset). 
The large-$c_0$  leading-order behaviour of $\cfk$ is the same as for $U>0$, shifted by $U$ compared with its value when $U=0$. 
Figure \fref{speed_small_scale_open_flow} (bottom row)
compares the two front speeds $\cfk$ and $\cg$.
Unlike the previous cases,
the difference and relative difference 
vary monotonically with $c_0$, with   peak values  as $c_0 \to -U$ when $\cg \to 0$ while $\cfk$ remains finite.

 \section{Conclusion}\slab{conc}
In this paper, we focus on the effect of spatially periodic flows on the propagation of the sharp chemical fronts that arise in the \eref{FK} model for small diffusion and fast reaction (large P\'eclet and Damk\"ohler numbers) and on their heuristic approximation by the \eref{G} equation. We introduce a variational formulation that expresses the long-time front speed in each model in terms of periodic trajectories minimising the time of travel across a period of the flow, thus providing  an alternative route to  the homogenization of the corresponding Hamilton--Jacobi equations. 
In this formulation, the difference between the front speeds predicted by the two models arises from a different constraint imposed on the minimising trajectories. This makes it easy to deduce that the  \eref{FK} front speed 
 is greater than or equal to the  \eref{G} front speed,  
 with equality in the case of shear flows.

We examine the front speed for a two-parameter family of periodic cellular flows in a channel, with both 
zero and non-zero mean velocity $U$, relying on  a numerical implementation of the variational representation. 
We find that for $U\geq 0$, the relative difference between the two front speeds is smaller than 10\% for a broad range of parameters with   the largest values obtained 
 when the reactions and mean flow are both relatively weak  ($\Da \gtrsim 1$ number and $U \ll 1$).
This is confirmed by the closed-form expressions we obtain in the two asymptotic limits $c_0=2 \sqrt{\Da/\Pe} \ll 1$ and $c_0 \gg 1$. 
For $U<0$, the relative difference between the two front speeds increases rapidly with decreasing $c_0$.  As $c_0 \to -U$, the \eref{G} front becomes stationary. There is then a range of $c_0<-U$ for which  right-propagating fronts exist for \eref{FK} but not  for \eref{G}. 
In this range \eref{G} fails completely as a heuristic model for \eref{FK} front, even at a qualitative level. The dramatic difference between the two models can be traced to the difference between the pointwise and time-integrated constraints that appear in the variational formulations  
\eref{cfkvar} and \eref{cgvar}.

A fundamental assumption that we make is that the minimising   trajectories 
that control the two  front speeds  inherit the  spatial periodicity of the background flow.  
We have carefully tested the validity of this assumption for the two-parameter
family of periodic cellular flows considered here  
against computations over domains of length twice and three times the  $2\pi$-period of the flow and found that the minimisers are $2\pi$ periodic. 
These results confirm that the front speed is indeed controlled by trajectories with the same periodicity as that of the flow. It would nonetheless be desirable to establish this property rigorously. A proof would also clarify whether it is specific to the class of flows considered here or holds more generally.

We have obtained the Hamilton--Jacobi \eref{I1} equation for \eref{FK} under the formal assumptions $\Pe \gg 1$, $\Da \gg 1$ and $\Da=O(\Pe)$ (so that $c_0=O(1)$). Its range of validity, and hence that of our results, is in fact much larger and includes small values of $\Da$. This is because it is only necessary for the WKBJ approximation leading to \eref{I1} to hold that $\Pe \nabla \mathscr{I}$ -- which involves a combination of $\Pe$ and $\Da$ -- be large. For shear flows, it follows from $\mathscr{I} = t \,\mathscr{G}(x/t,c_0) + O(1)$ and the form of $\mathscr{G}$ in \eref{Gshear} that the condition is satisfied provided that $\Da\gg \Pe^{-1}$, equivalent to the requirement that the front thickness in the absence of shear be small. The situation is more complex for cellular flows because of the logarithmic dependence that arise (see \eref{c1FKapp}). For standard cellular flows (with $A=U=0$), we can refer to \cite{TzellaVanneste2015} where the asymptotic of the front speed is derived for $\Pe \gg 1$ and arbitrary $\Da$, 
based on the computation of the principal eigenvalue
of the relevant advection--diffusion eigenvalue problem \cite{GartnerFreidlin1979,Freidlin1985,BerestyckiHamel2002}.  
It is found there that, as $\Da$ is reduced from large values, the Hamilton--Jacobi regime gives way to a different regime characterised by the scaling $\Da = (\log \Pe)^{-1}$ and requiring a delicate matched-asymptotics analysis. This indicates that the results of the present paper apply for $\Da \gg (\log \Pe)^{-1}$. The range of validity is presumably the same for $A \not=0$, but not for $U \not= 0$: in the latter case, since the small-$\Da$, i.e.\ small $c_0$ limit, is controlled by the flow around the (fastest) open streamlines, we expect the range of validity to be that of shear flows, that is, $\Da \gg \Pe^{-1}$. A complete analysis would require generalising the results of \cite{TzellaVanneste2015} to $U \not=0$, and to deal with the subtleties that arise in the limit $U \ll 1$ (cf.\ the effective-diffusivity computation in this regime in \cite{SowardChildress1990}).

We conclude by mentioning  three possible extensions of our work. 
The first concerns the shape of the front the \eref{FK} model, which can be determined from the solution to Hamilton--Jacobi equation \eref{I1}. Specifically, the front at time $T$ is 
the level curve 
 $\mathscr{I}(\bx,T,c_0)$, with
$\mathscr{I}(\bx,T,c_0)$   defined by the variational formula in \eref{I2}. In this case, the minimising trajectories are not periodic but satisfy the end condition $\bs{X}(T)=\bx$. For large $T$, they stay close to the periodic trajectories determining $\cfk$ for a long time interval before $T$, so the starting condition $\bs{X}(0)=(0,\cdot)$ can be replaced by a more practical condition that $\bs{X}(T-t)$ be asymptotic to the periodic trajectories as $t \to \infty$. 
The second extension concerns cellular flows in the entire plane, as opposed to the channel configuration considered in this paper. In this case, the problem is enriched by the two-dimensional nature of the front speed and the fact that 
minimising trajectories corresponding to speeds with irrationally related components cannot be periodic. 
Similarly, in the presence of a mean flow, the front speed is likely to depend sensitively on whether the two 
component of the flow velocity are 
rationally or irrationally related   
(the same is true for the components of the effective diffusivity tensor; see \cite{FannjiangPapanicolaou1994,MajdaKramer1999,PavliotisStuart2007}). 
It would be of interest to investigate how these aspects affect the differences between $\cfk$ and $\cg$. 
Finally, a third extension concerns other types of cellular flows. While $\cfk$ remains close to $\cg$  in the strong-flow regime $c_0 \ll 1$  
for the `cat's eye' flow 
(obtained by a periodic variation to the
basic cellular flow  \cite{ChildressSoward1989}),
the difference can become significant for  the (integrable) three-dimensional Roberts cellular flow  
\cite{XinYu2014}.
For more complex (non-integrable) flows, e.g.  
  the time-periodic, two-dimensional cellular flows 
  considered in \cite{CamassaWiggins1991}
  or the three-dimensional Arnold--Beltrami--Childress flows \cite{Dombre_etal1986},
  the situation is more challenging \cite{McMillen_etal2016}. 
These flows could be tackled by the analytic and numerical approaches employed in this paper. We leave this for future work.

\section*{Acknowledgments}
A. Tzella   gratefully acknowledges support from EPSRC (Grant No.\ EP/P511286/1).

\appendix

\section{Numerical procedure}\slab{num}
For \eref{FK}, we focus on the variational expression \eref{Glong} and approximate the periodic trajectory $\bX(t)$  
by a piecewise linear function $\bm{X}_d$, defined on an evenly spaced time grid $\{t_l=l\Delta t\}_{l=0}^N$ where $t_N=\tau$.  
The action functional in \eref{Glong} 
is approximated by the sum 
\beq
G_d(\{\bm{X}_l\}_{l=0}^N,c_0) =
\frac{1}{4}\left(\frac{1}{\tau}\sum_{l=0}^{N-1}L_d(\bm{X}_l,\bm{X}_{l+1}) -c_0^2\right), 
\eeq
where $\bm{X}_l=\bm{X}_d(l\Delta t)$ approximates $\bm{X}(t_l)$,
\beq
L_d(\bm{X}_l,\bm{X}_{l+1})=\Delta t \mathscr{L}\left(\frac{\bm{X}_{l+1}-\bm{X}_{l}}{\Delta t},\frac{\bm{X}_l+\bm{X}_{l+1}}{2}\right),
\eeq
with $\mathscr{L}$ is defined \eref{Lag}, and we have used a midpoint rule to approximate the integral. 
The symplectic nature of the midpoint rule (e.g.\ \cite{MarsdenWest2001}) ensures that the corresponding value of the Hamiltonian remains constant over time.

For \eref{G}, we focus on the variational expression \eref{cgvar2}.
Calculations are easiest taking $\Theta(x)$ to parameterise the pointwise constraint in polar coordinates yielding 
\begin{subequations} \elab{constraint2_G}
\begin{align}
	T'(x)&=\frac{1}{(u(x,Y(x))+c_0\cos\Theta(x))},\\
Y'(x)&=\frac{u(x,Y(x))+c_0\cos\Theta(x)}{v(x,Y(x))+c_0\sin\Theta(x)}, 
\end{align}
\end{subequations}
where $\Theta(x+2\pi)=\Theta(x)$.
We now approximate  $Y(x)$, $\Theta(x)$   and   $T(x)$ 
by piecesewise linear functions $Y_d$,  $\Theta_d$ and $T_d$, defined on an evenly spaced spatial grid $\{x_k=k\Delta x\}_{k=0}^N$ where 
$x_N=2\pi$. 
The total time period $\tau$ may then be approximated as  
\begin{subequations}\elab{discrete_G}
\beq
\tau_d(\{Y_k,\Theta_k\}_{k=0}^N)
=\sum_{k=0}^{N-1} T_{k+1}-T_k
\enspace\text{where}\enspace
 T_{k+1}-T_k\approx\int_{k\Delta x}^{(k+1)\Delta x}
T'(x)\d x
\eeq
subject to the constraint
\beq
Y_{k+1}-Y_k\approx \int_{k\Delta x}^{(k+1)\Delta x}Y'(x)\d x,\quad \textrm{for} \ \ k=1\ldots N. 
\eeq
\end{subequations}
Here, $Y_k=Y_d(k\Delta x)$, $\Theta_k=\Theta_d(k\Delta x)$ and $T_k=T_d(k\Delta x)$ are respectively an approximation to $Y(x_k)$, $\Theta(x_k)$ and $T(x_k)$.
 We use the midpoint rule to  approximate  the integrals in \eref{discrete_G} so that
\begin{subequations}\elab{discrete_G2}
\begin{align}
 T_{k+1}-T_k&=\Delta x
 \frac{1}{u\left(x_{k}+\frac{1}{2}\Delta x,\frac{1}{2}(Y_{k+1}+Y_{k})\right)+
 c_0\cos\left(\frac{1}{2}(\Theta_{k+1}+\Theta_{k})\right)}
 \\
 \intertext{and}
  Y_{k+1}-Y_k&=\Delta x\frac{
  u\left(x_{k}+\frac{1}{2}\Delta x,\frac{1}{2}(Y_{k+1}+Y_{k})\right)+
  c_0\cos\left(\frac{1}{2}(\Theta_{k+1}+\Theta_{k})\right)
  }
  {
    v\left(x_{k}+\frac{1}{2}\Delta x,\frac{1}{2}(Y_{k+1}+Y_{k})\right)+
    c_0\sin\left(\frac{1}{2}(\Theta_{k+1}+\Theta_{k})\right)  
    }.
\end{align}
  \end{subequations}

In both problems, we  use MATLAB's Symbolic Math Toolbox to express the trajectories, action functional and constraints in symbolic form. 
 We then 
 take $\Delta t=\tau/200$  and $\Delta x=\pi/100$ 
 and use MATLAB's Optimization
  Toolbox 
 to 
 find the optimal trajectories that
  minimise  the value of (i)  
 $T_d(\{Y_k,\Theta_k\}_{k=0}^N)$ 
 from where we obtain $\tg$ as a function of $c_0$
 and (ii)  $G_d(\{X_l,Y_l\}_{l=0}^N,c_0)$ from where we solve $\mathscr{G}(c,c_0)$. We then use \eref{Gc02} to deduce $c_0$ for a given $\cfk$.
 The advantage of symbolic calculations is that the gradient vectors of the discretised action functional and constraints 
 can readily be determined. 
 These are necessary to increase the accuracy and efficiency of the optimisation   solver.

The computations need a good first guess to be initialised.
For  problem \eref{cgvar}, we use the large-$c_0$ asymptotic behaviour of the trajectory obtained for the basic  
cellular flow with closed  streamlines ($A=U=0$) given by equation \eref{c2Gapp}.
We then iterate over a range of values of $c_0$ using the previously determined trajectory as an initial guess to find the next minimiser.
Similarly, for problem \eref{cfkvar} we use the large-$c$ asymptotic behaviour of the trajectory   given by equation \eref{c2FKapp}. 
The iteration is this time taking place over a range of values of $\cfk$.
The  same solutions are used  as  first guess to 
obtain the optimal solutions for a range of $A$ and $U$ values.

\bibliographystyle{siam.bst}
\bibliography{front}

\end{document}